\documentclass[conference]{IEEEtran}

\usepackage{cite}
\usepackage{amsmath,amssymb,amsfonts}
\usepackage{algorithmic}
\usepackage{graphicx}
\usepackage{textcomp}
\usepackage{bmpsize}
\usepackage{xcolor}
\usepackage{lipsum}
\usepackage{makecell} 
\usepackage{adjustbox}
\usepackage{enumitem}

\usepackage[capitalise,noabbrev]{cleveref}
\usepackage{tabularx}
\usepackage{booktabs}

\usepackage{url}

\usepackage[table]{xcolor}

\usepackage{subcaption}

\usepackage{siunitx}

\usepackage{xspace}
\newcommand{\name}{\textit{ThermalTap}\xspace}

\begin{document}

\title{ThermalTap: Passive Application Fingerprinting in VR Headsets via Thermal Side Channels}

\author{
    \IEEEauthorblockN{
        Mahsin Bin Akram, 
        A H M Nazmus Sakib, 
        OFM Riaz Rahman Aranya, \\
        Raveen Wijewickrama, 
        Kevin Desai, and 
        Murtuza Jadliwala
    }
    \IEEEauthorblockA{
        \textit{University of Texas at San Antonio}\\
        San Antonio, Texas, USA \\
        \{mahsin.akram, ahmnazmus.sakib\}@my.utsa.edu, \\
        \{ofmriazrahman.aranya, raveen.wijewickrama, kevin.desai, murtuza.jadliwala\}@utsa.edu
    }
}

\maketitle

\begin{abstract}

Standalone virtual reality (VR) headsets process highly sensitive personal, professional, and health-related data, yet their susceptibility to non-contact physical side channels remains largely unexplored. Existing side-channel attacks typically require malicious software execution or physical access to peripherals, making them conspicuous and potentially patchable. This paper introduces ThermalTap, the first passive, non-contact side-channel attack that fingerprints VR applications solely from the long-wave infrared (LWIR) radiation emitted by the headset chassis. 
By treating a headset's thermal signature as a high-fidelity proxy for internal computational workloads, ThermalTap enables remote application inference at meter-scale distances without any device interaction. To achieve robust performance in real-world settings, the system combines a commodity thermal camera with a multi-modal sensor suite (capturing ambient temperature, humidity, and airflow) to normalize environmental noise. We evaluate ThermalTap using six applications across three commercial standalone headsets. 
In indoor settings, ThermalTap identifies applications with over 90\% accuracy using only 10 seconds of thermal camera data. Under outdoor conditions, with longer session-level observations, several applications remain identifiable despite environmental variability, with the strongest outdoor application reaching 81\% accuracy.
Our findings establish thermal radiation as a fundamental and unavoidable privacy risk for immersive systems, exposing a critical security gap that bypasses current software-level protections and physical access controls.

\end{abstract}

\section{Introduction}
\label{sec:introduction}
The proliferation of standalone virtual reality (VR) headsets has transformed them from niche entertainment peripherals into general-purpose computing platforms. Devices such as the Meta Quest 2, Meta Quest 3, and HTC Vive Focus Vision are increasingly deployed across enterprise, healthcare, and education domains, where they process deeply personal information including health and wellness data, professional communications, and intimate social interactions~\cite{zhang2023s,li2024dangers}. Unlike traditional computing devices, VR headsets are worn on the face, operate autonomously without a tethered host, and are often used in shared or semi-public environments such as offices, university labs, and VR arcades. This combination of sensitive data, physical co-location of users, and a form factor that inhibits user situational awareness creates a compelling and underexplored attack surface.

Central to this threat is \textit{application fingerprinting}: identifying which VR application a victim is currently running. This inference may appear coarse, but it can reveal sensitive contextual information about the user's immediate activity. A mental health or wellness application may reveal therapeutic or health-related behavior; an enterprise collaboration tool may disclose professional activity or organizational affiliation; a social VR application may reveal communication or social interaction; and a training or simulation application may reveal institutional or operational context. Repeated observations can further expose activity timelines, such as when a user enters a meeting, starts a training session, or switches between entertainment and work applications. Such leakage can support behavioral profiling, context-aware social engineering, targeted surveillance, and organizational intelligence gathering, even when the adversary never observes the headset display or interacts with the device \cite{Son2025SidechannelIO, Siddens2026MemoryDM, zhang2023s}.

Prior work has shown that VR activity leaks through both software-accessible and physical side channels~\cite{zhang2023s,li2024dangers,Sun2025VReavesEO}. Software-accessible runtime signals, such as rendering and performance information exposed by AR/VR platforms, can reveal user interactions and identify applications launched by a victim~\cite{zhang2023s}. Physical channels provide another leakage path: cable-based measurements can reveal visual- and audio-related VR activity during charging~\cite{li2024dangers}. Despite their effectiveness, existing VR side-channel attacks share a fundamental constraint: they require either software access to the victim's headset or physical access to a device peripheral. These requirements increase attacker effort and make the intrusion easier to detect via software patches or physical inspection.

We observe that a more fundamental and unavoidable information channel has remained unexamined in the VR context: \textit{thermal emission}. Every electronic device dissipates heat as a direct consequence of electrical current flowing through resistive components. 
In the context of VR, different applications impose distinct, high-intensity workloads on the processor, display pipeline, and cooling systems.
This heat propagates to the device's external surface and radiates into the surrounding environment as long-wave infrared (LWIR) radiation, which is detectable by commercial thermal cameras at a distance. Our key insight is that these thermal signatures are a high-fidelity proxy for the internal workload, and because heat dissipates spatially across the chassis, it creates a unique ``thermal fingerprint'' for specific applications.

In this paper, we introduce \textbf{\name}, a passive, non-contact thermal side-channel attack that fingerprints VR applications from radiometric observations of the headset chassis. An attacker equipped with a commodity infrared thermal camera can stealthily observe a victim wearing a VR headset from across a room and infer the active application without software access, cable instrumentation, or interaction with the victim device.
However, turning external thermal radiation into a usable side channel requires solving three measurement problems: capturing absolute per-pixel temperature rather than colorized video, isolating the headset surface from the user's face and surrounding background despite thermal blooming under load, and accounting for environmental confounds (ambient temperature, airflow, distance) that alter observed surface temperature without being caused by the running application. \name addresses these through thermal radiometric capture, learned headset-region segmentation, spatiotemporal feature extraction over the segmented region, and synchronized environmental sensing, with supervised classification mapping the resulting thermal signature to an application label.

We evaluate \name using six active VR applications across indoor and outdoor settings spanning different ambient temperatures, airflow, and solar exposure conditions.
In indoor experiments, \name correctly identifies active applications with approximately 90\% accuracy using only \SI{10}{second} thermal observation windows. Outdoor conditions reduce overall performance due to environmental variability, but longer session-level observations remain effective for several applications, with the strongest outdoor case reaching an accuracy of 81\%.
These results show that externally emitted headset heat contains app-specific information that can be recovered from short, passive observations of VR devices.

In summary, this paper makes the following contributions:

\begin{itemize}[leftmargin=*]
    \item \textbf{A novel attack vector:} We are the first to demonstrate external, non-contact thermal side-channel attacks on VR headsets, establishing thermal emission as a practical and stealthy information leakage channel that requires no software or hardware access to the victim's device.

    \item \textbf{Spatiotemporal Fingerprinting:} We propose a novel representation that partitions the VR headset chassis into a spatial grid, capturing the temporal evolution of heat to distinguish between apps with similar instantaneous temperatures.
    
    \item \textbf{End-to-End System:} We design and implement a complete attack system comprising a custom multi-modal Raspberry Pi-based sensor apparatus, an automated ROI segmentation pipeline, and an application fingerprinting classifier.

    \item \textbf{Comprehensive evaluation:} We evaluate \name with six VR applications and three commercial standalone headsets across indoor and outdoor environments and cross-device settings. We also study practical impact factors including required attack thermal camera distance, observation durations, environmental variation, and device variability.
    
\end{itemize}

\section{Background and Related Works}
\label{sec:background}

This section positions \name within prior work on physical side channels and VR security, and explains why external thermal emissions represent a distinct attack surface.

\noindent\textbf{Side-Channel Attacks and VR Threats}
Side-channel attacks exploit unintended physical or software-observable signals to infer sensitive information about a system's internal activity. Prior work has shown that internal system activity leaks through multiple modalities, including power consumption \cite{genkin2014powersidechannel}, electromagnetic emanations \cite{hayashi2014threat}, radio-frequency signals \cite{ni2023eavesdropping}, and acoustic emissions \cite{luo2024eavesdropping}. For example, power-side channels have been used to extract cryptographic secrets and infer smartphone activity through charging lines~\cite{Cronin2021ChargerSurfingEA}, while EM and RF-based attacks have been used to reconstruct device activity or classify mobile applications~\cite{chen2024eavesdropping,hayashi2014threat,ni2023eavesdropping}. Acoustic side channels have also been used to infer user input, including keystrokes and VR controller interactions~\cite{luo2024eavesdropping,Ayati2025MakingAS}.

VR systems are especially attractive targets for side-channel attacks because they are self-contained computing platforms with rich sensors, sustained and heterogeneous compute workloads, and sensitive user interactions. Existing AR/VR attacks have exploited software-accessible signals, sensors, peripherals, and wireless channels. Zhang et al.~\cite{zhang2023s} showed that a malicious background application can use exposed performance and rendering information to infer user interactions and fingerprint concurrently running applications. LineTalker~\cite{li2024dangers} demonstrated that VR application activity leaks through charging-cable power measurements. Other attacks infer user input/activity through hand-tracking cameras, acoustic signals, WiFi channel state information~\cite{ni2024non,cayir2025speak,Arafat2021VRSpyAS,He2025AcouListenerAI,Sun2025VReavesEO,yang2024keystroke}. Among these, the closest physical side-channel attack to our work is VReaves~\cite{Sun2025VReavesEO}, which fingerprints VR applications from electromagnetic emanations using a software-defined radio. While both attacks require no software access to the victim, they exploit fundamentally different signals. VReaves relies on instantaneous, high-frequency EM leakage from hardware components such as clock lines and buses, and requires specialized, relatively costly radio equipment and signal processing. In contrast, \name aims to capture the time-integrated thermal footprint of the entire device using a commodity infrared camera. Further, EM leakage may be mitigated by a simple randomization of computational patterns, which directly alters high-frequency switching behavior. In contrast, while such randomization can perturb thermal signals, thermal emissions reflect the time-integrated power of the whole device and are therefore harder to obfuscate without incurring substantial overhead or requiring hardware-level changes.

In contrast, \name exploits a fundamentally different and more operationally feasible leakage channel: the passive infrared radiation emitted by the headset chassis itself, observable at meter-scale distances using only a low-cost commodity thermal camera, with no software access, cable instrumentation, or specialized equipment required

\noindent\textbf{Thermal Side Channels}
Thermal leakage arises because computational activity consumes power and produces heat. As heat propagates through the device and reaches the external chassis, it creates a spatial and temporal surface-temperature pattern. Different applications can induce different workload patterns across the processor, display pipeline, memory, networking, and cooling system, causing the externally visible thermal profile to vary over time.
Prior thermal side-channel work has primarily focused on residual heat from user input surfaces or internal device sensors. Thermal imaging has been used to recover keyboard or keypad input from residual heat traces~\cite{Wodo2016ThermalIA,Mowery2011HeatOT,Abdrabou2020AreTA,Macdonald2023ConductingAM,Alotaibi2022ThermoSecureIT,Abdelrahman2017StayCU}. In contrast, ThermalBleed showed that software-accessible internal thermal sensors on x86 CPUs can leak microarchitectural state and break KASLR~\cite{kim2022thermalbleed}.

These works establish thermal behavior as an information-bearing channel, but they do not address external, non-contact thermal observation of VR headsets. Unlike keyboards, a VR headset continuously dissipates heat from sustained application workloads. Unlike internal sensor attacks, \name does not require software access to the victim device. This gap motivates our study of whether externally radiated thermal signatures from consumer VR headsets can be used for application fingerprinting.

\section{Threat Model \& Research Questions}
\label{sec:threat}

\subsection{Threat Model}
\label{subsec:threat_model}
\name performs passive application fingerprinting from externally observable thermal emissions. The attacker's goal is to infer which VR application a victim is currently running, without directly interacting with the victim, the headset, or any headset peripheral.

\begin{figure}[htb]
    \centering
    \includegraphics[width=0.99\linewidth]{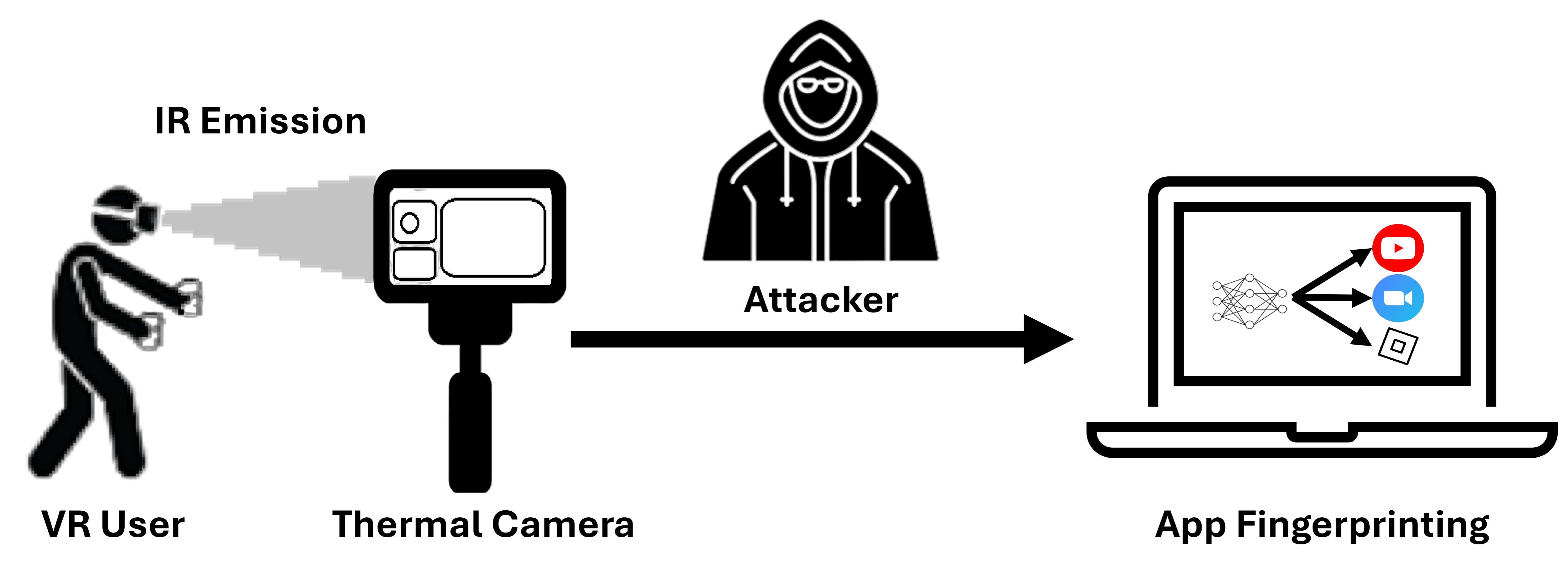}
    \caption{Adversary model.}
    \label{fig:adv_model}
\end{figure}

\noindent\textbf{Attack Setting}
We consider an adversary equipped with a commodity thermal camera, such as an Infiray P2 Pro or Topdon TC001, connected to a small computing device for data collection and inference. The camera is surreptitiously placed with line-of-sight to the victim's headset and passively records long-wave infrared emanations from the headset chassis.
This setting is plausible in shared or semi-public environments where VR users and bystanders may remain co-located for extended periods, such as university labs, shared offices, dorm common rooms, VR arcades, and training facilities. The thermal camera can be placed in ordinary room infrastructure or carried by a nearby observer, and the victim's situational awareness is naturally reduced while wearing the headset. Similar proximity-based observation models have been considered in prior side-channel attacks~\cite{chen2024eavesdropping,hayashi2014threat,luo2024eavesdropping}.

We assume the adversary can observe or infer the headset model during data collection. This is realistic because the device is visible to the observer and because commercial headset models have distinct external form factors. This assumption is also important for cross-device evaluation, since our results show that device-specific thermal behavior affects transferability.

\noindent\textbf{Adversary Capabilities and Scope}
We assume the adversary has the following capabilities and limitations:

\begin{itemize}[leftmargin=*]
    \item The adversary has no physical access to the victim's headset or its peripherals (hand controllers, charging cables, link boxes, companion mobile devices).
    \item The adversary has no software-level access to the headset, does not install a malicious application, and does not rely on permissions, APIs, or internal sensors exposed by the VR platform.
    \item The adversary performs only passive, non-contact observation. \name does not transmit signals, inject workloads, or interact with the victim device.
    \item The adversary possesses training data for a predefined set of candidate VR applications. When device-specific modeling is required, the adversary may collect training data from the same headset model as the victim.
    \item The adversary's objective is application fingerprinting over a known candidate set rather than open-world application discovery. This is consistent with the deployment reality of standalone VR platforms: each platform exposes a curated store with a finite catalog, and a small set of popular applications dominates real-world usage. An adversary can therefore profile this candidate set offline before deployment.
\end{itemize}

Inferring the application currently being run by the victim user can reveal sensitive contextual information about the user, such as whether they are engaged in entertainment, communication, work, training, health, or wellness-related activities. However, \name does not attempt to infer fine-grained in-app actions, typed input, biometric attributes, or user identity; these are outside the scope of this work.

\subsection{Research Questions}
\label{sec:rqs}

Our evaluation addresses three research questions that guide the design and analysis of \name. Together, they characterize the thermal side channel along three dimensions: whether external thermal emissions reveal application-level information, whether this signal remains useful under variable environmental conditions, and how headset-specific hardware characteristics affect generalization.

\noindent\textbf{RQ1 (Application Fingerprinting):} 
\emph{Can \name infer currently executing VR application from externally observable thermal emissions?}

\noindent This RQ evaluates the core leakage claim of \name. We examine whether different VR applications induce distinguishable thermal signatures on the headset chassis, and whether these signatures can be used to classify the currently executing applications from infrared measurements captured using commodity thermal cameras.

\noindent\textbf{RQ2 (Environmental Robustness):} 
\emph{How robust is \name under outdoor environmental conditions?}
\noindent This RQ evaluates whether application-level thermal signatures remain useful when the observation setting itself introduces variation unrelated to the running application. In particular, outdoor scenarios introduce unpredictable conditions, including ambient temperature changes, natural airflow, and solar-induced heating, all of which can alter the observed thermal signal independently of the application workload.

\noindent\textbf{RQ3 (Cross-Device Effects):}
\emph{To what extent do headset-specific hardware characteristics affect thermal application fingerprinting?}
\noindent This RQ evaluates whether the thermal signatures used by \name are primarily app-specific or strongly shaped by the VR hardware on which the application runs. Commercial standalone headsets differ in SoC placement, battery location, cooling design, chassis geometry, and material properties, all of which can affect how application-induced heat propagates to the externally visible surface. Since the adversary has line-of-sight to the victim, the headset model can often be visually identified and profiled in advance. However, if thermal signatures transfer across headset models, the attacker would require less device-specific training data.

\section{Initial Observations \& Design Requirements}
\label{sec:design}

External thermal emissions are only useful for fingerprinting if application-induced heat remains distinguishable after propagating through the headset chassis and being observed externally. We therefore begin with a controlled pilot study to test whether this signal is measurable, identify which aspects of the thermal trace are informative, and derive the measurement requirements that shape \name.

\subsection{Pilot Setup}
\label{subsec:pilot_setup}

We instrumented a Meta Quest 2 with a TopDon TC001 thermal camera~\cite{meta_quest_2_2020,topdon_tc001_2022} positioned approximately \qty{60}{\cm} from the headset chassis with line-of-sight to the front face. Camera distance, ambient temperature, and wearer pose are held approximately fixed across trials, so the only deliberately varied factor is the executing application. The pilot suite spanned idle, media, utility, and simulation workloads (see in \cref{tab:app_list_feasibility}). 
Across 16 sessions we collected approximately 120 minutes of synchronized radiometric thermal recordings.
From each sample we extracted a preliminary feature set comprising global temperature statistics (mean, max, standard deviation, entropy), temporal features describing how the surface temperature evolves over time (thermal drift), and contextual metadata (camera-to-headset distance, ambient temperature). We then trained pilot a Random Forest classifier on this feature set to make our preliminary observations.

\begin{figure}[htb]
  \centering
  \begin{subfigure}[b]{0.4\columnwidth}
    \centering
    \includegraphics[width=\textwidth]{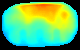}
    \caption{YouTube}
    \label{fig:thermal_yt}
  \end{subfigure}
  \begin{subfigure}[b]{0.4\columnwidth}
    \centering
    \includegraphics[width=\textwidth]{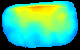}
    \caption{VRFS}
    \label{fig:thermal_vrfs}
  \end{subfigure}
  \caption{Thermal traces from two apps on a Meta Quest 2.} 
  \label{fig:thermal_comparison}
\end{figure}

\subsection{Observations}
\label{subsec:pilot_observations}
\begin{figure}[htb]
    \centering
    \includegraphics[width=0.8\linewidth]{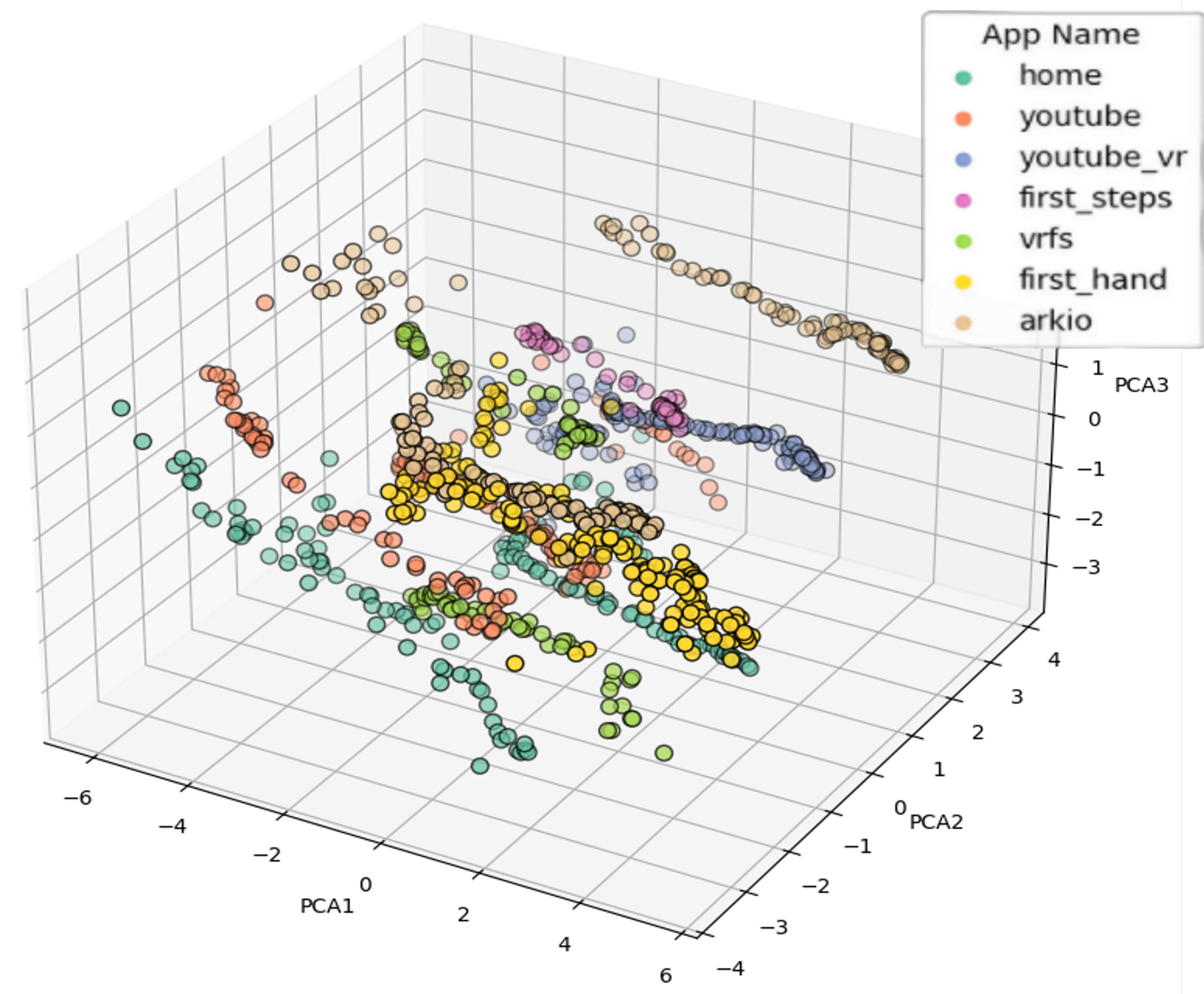}
    \caption{PCA projection of thermal feature vectors.} 
    \label{fig:feasibility_pca}
\end{figure}

\noindent\textbf{O1: Application-level thermal behavior is visibly separable.}
\Cref{fig:thermal_comparison} shows thermal frames for two pilot applications. VRFS exhibits a concentrated central hotspot near the SoC, while YouTube produces a more diffuse pattern dominated by display-backlight heating. We further project the preliminary feature vector described above via PCA (\cref{fig:feasibility_pca}), where application classes form visibly separable clusters along the first two principal components. This indicates that the externally observed thermal channel carries application-discriminative information. 

\begin{figure}[htb]
    \centering
    \includegraphics[width=1\linewidth]{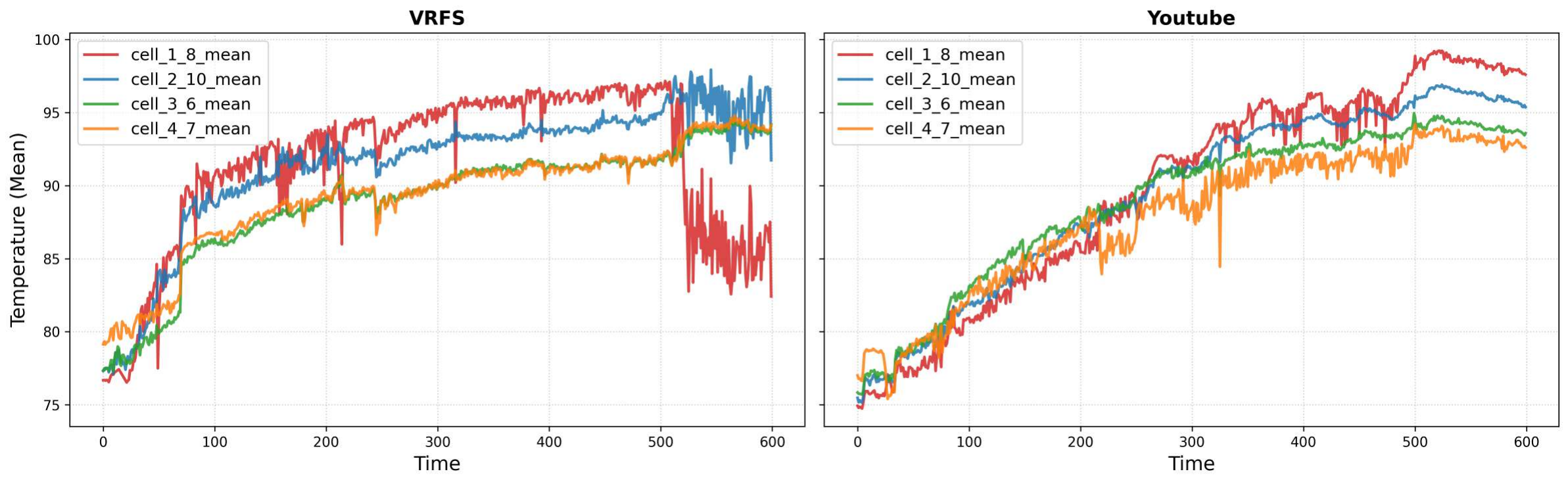}
    \caption{Thermal drift over time for four representative grid cells across the application sessions for VRFS (left) and Youtube (right).} 
    \label{fig:pilot_thermal_drift}
\end{figure}

\noindent\textbf{O2: Spatiotemporal evolution carries substantial application signal.}
A single thermal frame captures the headset chassis at one instant, but application-induced heating is inherently both spatial and temporal.
\textit{Spatially}, heat is not distributed uniformly across the chassis. Different workloads concentrate heat over different internal components, so applications can differ not only in average surface temperature but also in where heat appears. This motivates preserving the local spatial structure of the headset.
\textit{Temporally}, the surface temperature is not static. Different applications affect how quickly local regions heat up and how far they move away from ambient temperature. We refer to this temperature-versus-time behavior, especially its local rate of change of temperature, as \emph{thermal drift}. \Cref{fig:pilot_thermal_drift} illustrates this effect for apps VRFS and Youtube.

\begin{figure}[htb]
    \centering
    \includegraphics[width=0.72\linewidth]{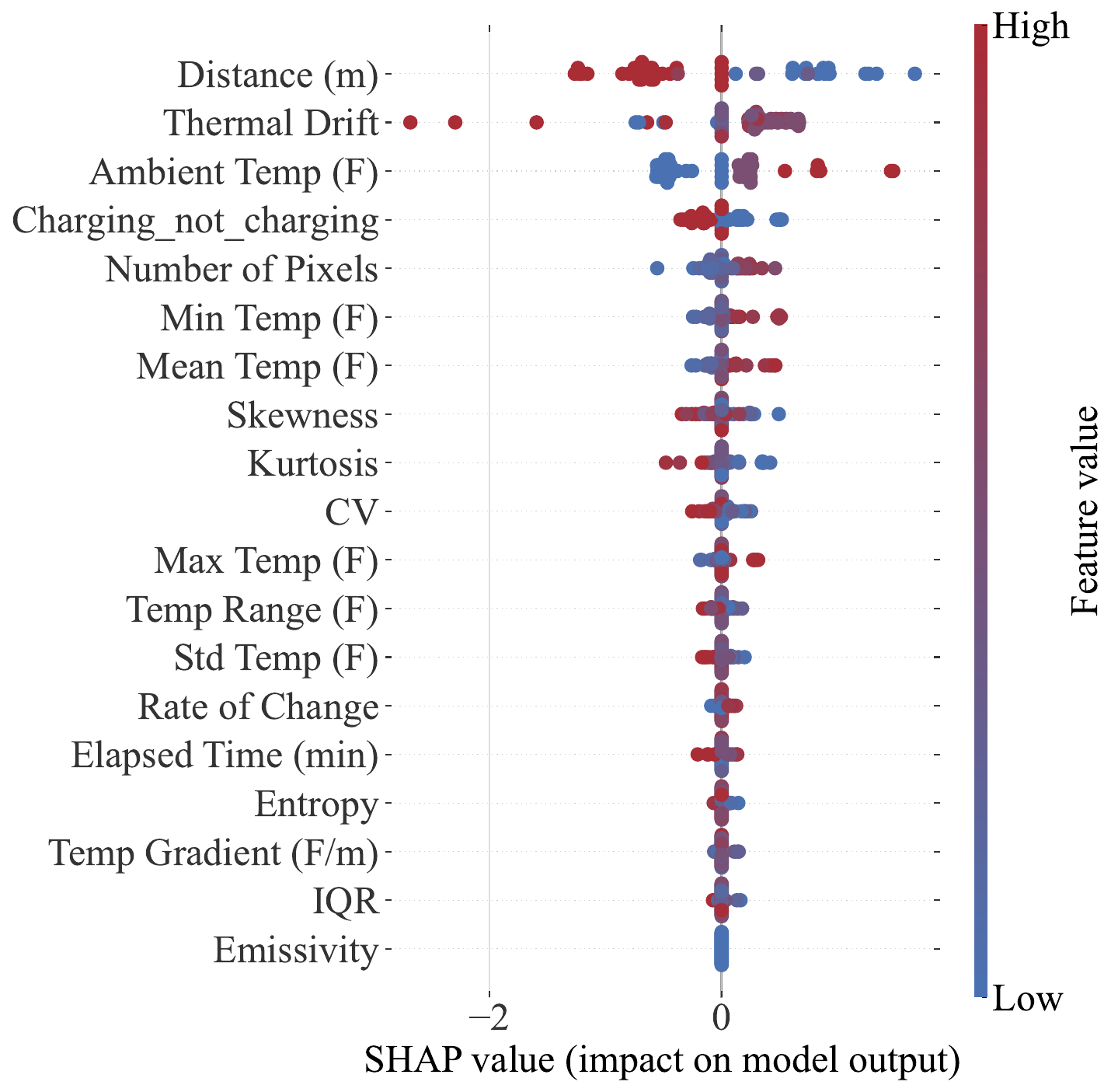}
    \caption{SHAP feature-impact analysis.}
    \label{fig:feasibility_shap_analysis}
\end{figure}

This is further supported by the SHAP analysis in \cref{fig:feasibility_shap_analysis}, where thermal drift emerges as one of the most influential features (apart from camera-to-headset distance and ambient temperature). 

\noindent\textbf{O3: Reliable fingerprinting requires isolating the headset region.}
The pilot showed that application-discriminative thermal information is concentrated on the headset chassis, but the raw thermal frame also contains the wearer's face, straps, and surrounding background. This makes naive whole-frame analysis unreliable, since non-headset regions can introduce temperature patterns unrelated to the running application. The problem is further complicated by natural head movement: as the wearer shifts pose, the headset's position, scale, and visible boundary can change across frames. A simple thermal-contrast approach is also insufficient: during high-load applications, heat may bloom across the headset surface and blur the boundary between the chassis and nearby regions, causing contour-based masks to either omit parts of the headset or include non-headset pixels. This shows the need for a reliable Region of Interest (ROI) extraction step that consistently isolates the headset surface before downstream analysis.

Together, these observations translate into three design requirements for \name. The system must: (1) extract a reliable headset ROI before analysis; (2) represent the headset as a spatiotemporal thermal signature, using radiometric temperature values to preserve both where heat appears and how it changes over time; and (3) record measurement context, such as ambient temperature and camera-to-headset distance, so non-application effects can be accounted for during analysis.

\begin{figure}[htb]
    \centering
    \includegraphics[width=0.9\linewidth]{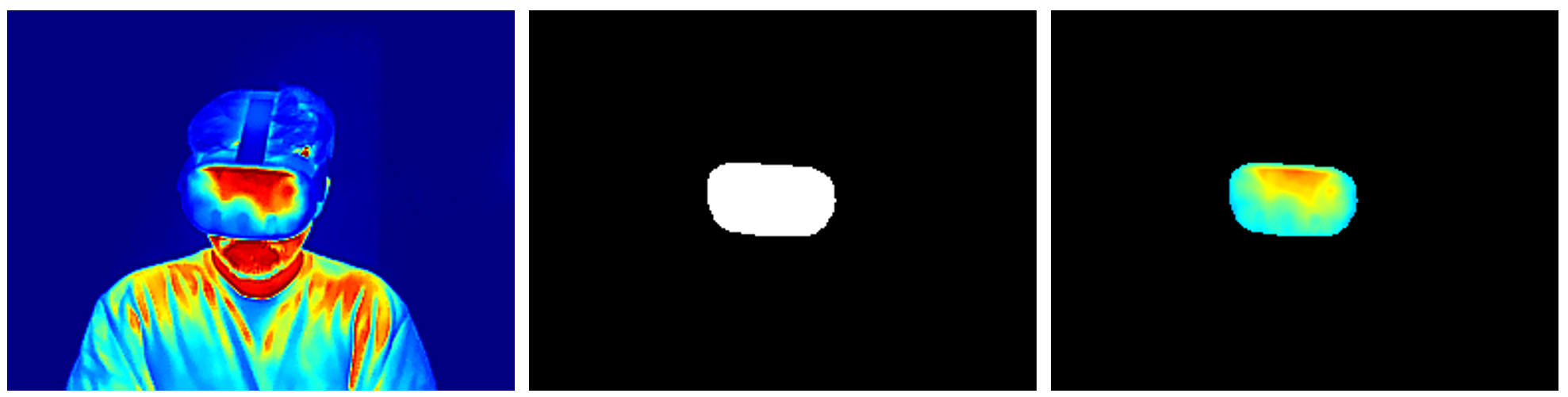}
    \caption{Headset ROI segmentation.} 
    \label{fig:segmentation}
\end{figure}

\section{\name Attack Framework}
\label{subsec:system_overview}

\begin{figure*}[htb]
    \centering
    \includegraphics[width=0.9\linewidth]{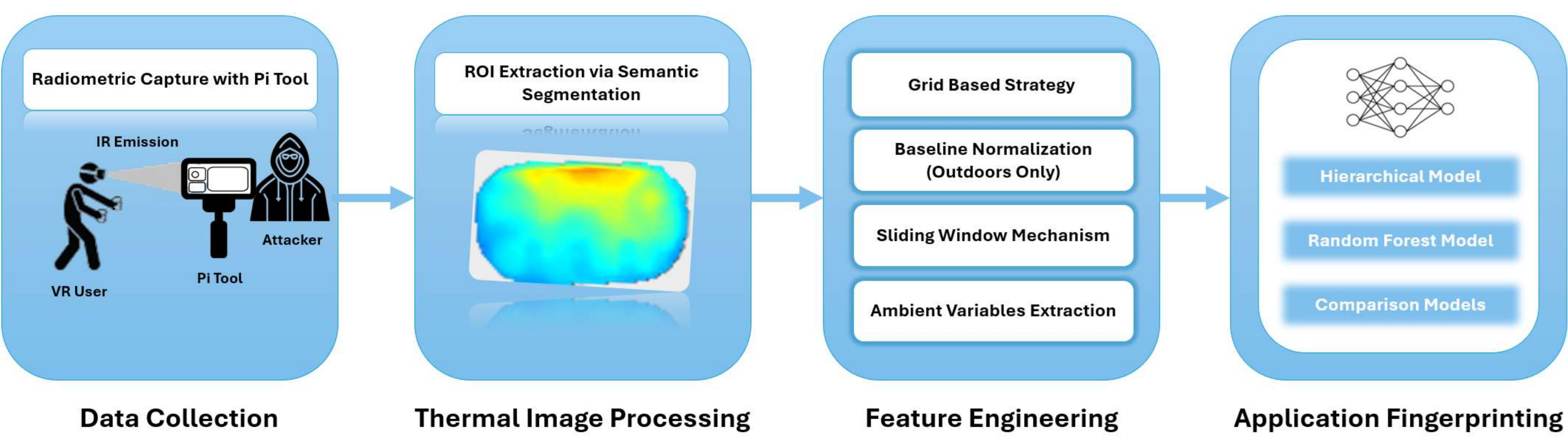}
    \caption{\name attack framework.}
    \label{fig:thermoscope_overview}
\end{figure*}

\cref{fig:thermoscope_overview} shows the \name pipeline. The acquisition thermal camera hardware apparatus (see \cref{subsec:apparatus} for details) records thermal frames from the headset surface together with ambient temperature, humidity, airflow, and camera-to-headset distance. The preprocessing module segments the headset region of interest (ROI) and removes unreliable masks. The feature extraction module converts each segmented frame into a spatial grid representation and augments it with temporal derivatives. Finally, the fingerprinting module maps each fixed-duration observation window to an application label.

We use three temporal units throughout the paper. A \emph{frame} is one timestamped radiometric measurement of the headset surface, stored as a matrix of per-pixel temperatures. A \emph{prediction window} is a fixed-duration contiguous sequence of frames from the same recording; \name produces one window-level prediction per window. A \emph{session} is one uninterrupted recording of a VR application from launch to exit; session-level predictions are obtained by majority vote over the window-level predictions in the session.

\subsection{Headset ROI Segmentation}
\label{subsec:roi_segmentation}
A raw thermal frame contains the headset chassis, the wearer's face, straps, and surrounding background. A natural first approach is to exploit thermal contrast, since the headset often appears warmer than the background. However, our initial experiments showed that thresholding, contour extraction, and generic segmentation were unreliable under high-load applications: heat blooms across the chassis and weakens the visual boundary between the headset, skin, straps, and nearby warm surfaces. This can either exclude heated headset regions or include non-headset pixels, directly contaminating the thermal features.
To address this, we developed a custom segmentation pipeline based on SRA-Seg~\cite{sraseg}, a semi-supervised segmentation method traditionally used for medical image segmentation. Training a segmentation model requires manually annotated headset masks, but densely labeling tens of thousands of thermal frames is labor-intensive. We therefore annotate only a subset of frames and use SRA-Seg to leverage the remaining unlabeled recordings during training. \Cref{fig:segmentation} shows an example of the resulting headset ROI segmentation. Further model and training specific details are provided in \cref{subsec:segmentation_training}.

To reject malformed masks, we apply geometric filtering to the largest connected component in each predicted mask. A valid headset mask must satisfy two constraints: a width-to-height ratio between 1.3 and 2.8, matching the front-facing shape of consumer headsets, and a solidity of at least 0.80, where solidity is the ratio of mask area to convex-hull area. These checks reject masks that include straps, face regions, or background artifacts. Frames failing either check are flagged as missing observations, preserving temporal alignment without contaminating the feature representation.

\subsection{Thermal Signature Representation}
\label{subsec:thermal_signature}

Motivated by O2 in \cref{subsec:pilot_observations}, \name represents thermal leakage as a spatiotemporal signature. Specifically, we define a \emph{thermal signature} as a time-series of per-frame feature vectors extracted from the segmented headset region. For an observation window of length $W$ seconds, this is $S_w = \{f_{t_0}, f_{t_0+1}, \ldots, f_{t_0+W-1}\},$ where each $f_t$ summarizes the \emph{spatial} structure of headset surface temperature at time $t$.
To construct $f_t$, \name partitions the headset region into an $N\times N$ grid. The resulting cells are used to compute two types of frame-level features: per-cell temperature statistics and per-cell spatial gradients.

\begin{figure}[h]
\centering
    \begin{subfigure}{0.4\linewidth}
      \centering
      \includegraphics[width=\linewidth]{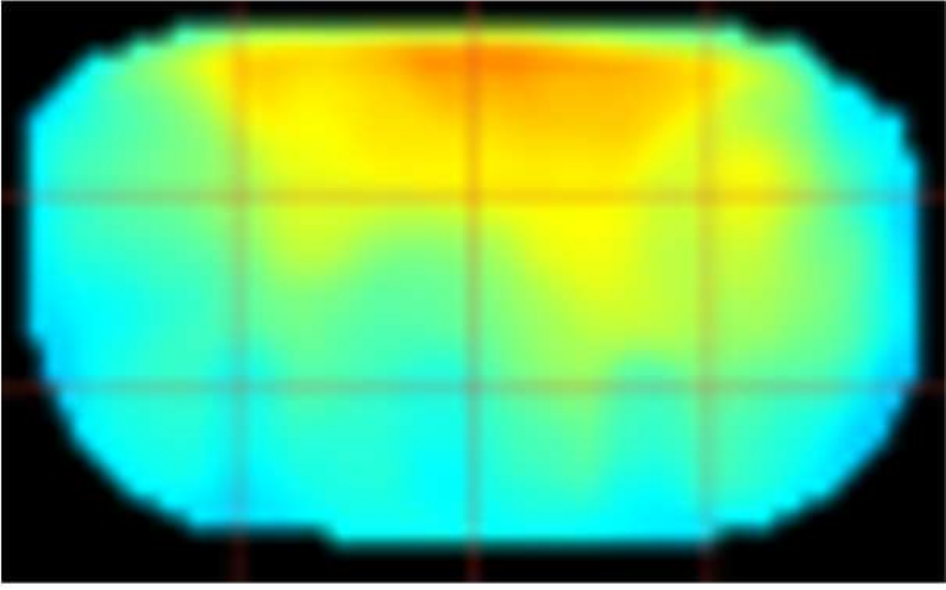}
      \caption{4x4 Grid Layout}
      \label{fig:sub1}
    \end{subfigure}
    \begin{subfigure}{0.4\linewidth}
      \centering
      \includegraphics[width=\linewidth]{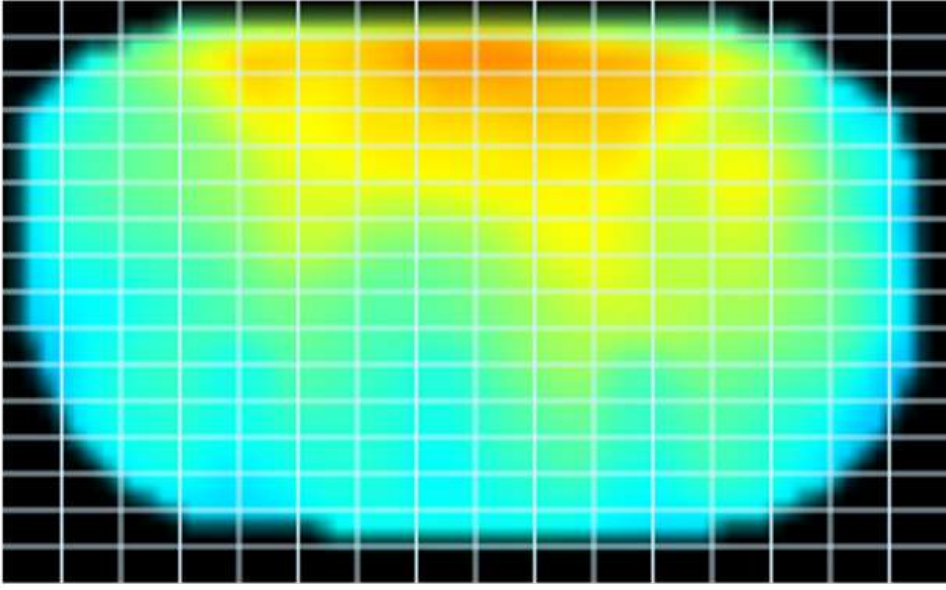}
      \caption{16x16 Grid Layout}
      \label{fig:sub2}
    \end{subfigure}
\caption{Grid layout on generated masks.}
\label{fig:grid_layout_masks}
\end{figure}

\noindent\textbf{Grid Resolution.}
Grid resolution $N$ controls the trade-off between noise robustness and localized thermal detail: coarse grids average over larger regions and suppress noise, while fine grids preserve localized hotspots but increase feature dimensionality and sensitivity to segmentation artifacts (see \cref{fig:grid_layout_masks} for a sample grid). We evaluate $N\in\{4,8,12,16,20,24\}$ in \cref{subsec:e1_results}.

\noindent\textbf{Per-cell statistics.}
For each segmented frame, \name partitions the headset bounding box into an $N\times N$ grid. From the valid headset pixels in each cell $(i,j)$ we extract four statistics: $\min_{i,j}(t)$, $\max_{i,j}(t)$, mean $\mu_{i,j}(t)$, and standard deviation $\sigma_{i,j}(t)$. A cell is considered valid only when at least 50\% of its area overlaps the headset mask; cells below this threshold are discarded to prevent background contamination. This yields up to $4N^2$ base features per frame.

\noindent\textbf{Spatial gradients.}
For each cell, we compute the difference between its mean temperature and the average of its cardinal neighbors (up, down, left, right):
\[
\mathrm{grad}_{i,j}(t) = \mu_{i,j}(t) - \frac{1}{|N_{i,j}|}\sum_{(k,l)\in N_{i,j}} \mu_{k,l}(t),
\]
where $N_{i,j}$ is the set of valid cardinal neighbors. This captures localized hotspots relative to surrounding regions independent of absolute temperature.

\subsection{Environmental Normalization}
\label{subsec:env_normalization}
During our initial pilot study, we observe that thermal measurements collected in highly variable conditions such as outdoor environments are affected not only by application-induced heating, but also by environmental factors such as ambient temperature and airflow. To reduce these effects, \name incorporates readings from our auxiliary environmental sensors on the raw thermal measurements.

\noindent\textbf{Baseline-derived temporal-delta subtraction.}
For each $(\text{application}, N\times N \text{cell})$ pair, we estimate from the indoor baseline training data the typical short- and medium-term temperature evolution of the chassis under that workload, expressed as per-cell delta profiles at two lag horizons:
\[
\Delta^{(\ell)}_{i,j}(t) = \mu_{i,j}(t) - \mu_{i,j}(t-\ell), \quad \ell\in\{5,30\}~\text{s}.
\]
Here, \(\Delta^{(\ell)}_{i,j}(t)\) is the temperature change of cell \((i,j)\) over the previous \(\ell\) seconds, \(t\) is the current timestamp, and \(t-\ell\) refers to the timestamp \(\ell\) seconds earlier. $\mu_{i,j}(t)$ is the mean temperature of the cell ${i,j}$ at time $t$.

Since thermal frames are sampled at \SI{1}{hz}, the \SI{5}{s} and \SI{30}{s} deltas compare the current cell temperature against the same cell 5 and 30 frames earlier, respectively.
At outdoor inference time, the corresponding indoor delta profile is subtracted from the observed outdoor delta, leaving a residual that reflects environmental and device variation rather than application workload. 

\noindent\textbf{Ambient and airflow correction.}
We also compensate for ambient temperature (collected by the BME280 sensor) by subtracting the measured ambient temperature from each grid-cell temperature value. This step shifts the representation from absolute surface temperature toward application-induced thermal load:
\[
L_{i,j}(t) = T_{i,j}(t) - T_{\mathrm{amb}}(t),
\]
where \(T_{i,j}(t)\) denotes the temperature of grid cell \((i,j)\) at time \(t\), and \(T_{\mathrm{amb}}(t)\) denotes the corresponding ambient temperature measurement.

Finally, we compensate for convective cooling effects (caused by wind) by adjusting temporal change features using measured air velocity (FS3000-1005 airflow sensor):
\[
\Delta T'_{i,j}(t) =
\left(T_{i,j}(t) - T_{i,j}(t-\Delta t)\right)\cdot (1 + k\,v(t)),
\]
where \(v(t)\) is the measured air velocity and \(k\) is an empirically estimated wind-compensation coefficient.

\subsection{Application Inference}
\label{sec:classification}

The VR headset's Home state represents an idle system state rather than a target application workload, \name therefore uses a two-stage inference procedure.

\noindent\textbf{Stage 1: activity-state detection.}
\name first determines whether the window corresponds to idle headset behavior or active application use. This stage uses a binary classifier trained to distinguish the Home state from all active-application states. If the majority of window-level predictions within the session indicate idle behavior, \name outputs the idle state and stops. Otherwise, the session is passed to the application fingerprinting stage.

\noindent\textbf{Stage 2: active-application fingerprinting.}
For sessions classified as active, \name applies a multi-class application classifier trained only on active VR applications. The classifier predicts an application label for each window in the session, and \name assigns the session to the application receiving the majority of window-level predictions. This aggregation reduces sensitivity to transient window-level errors while allowing \name to make predictions from short captured segments.

Both stages use the same spatiotemporal feature representation described in \cref{subsec:thermal_signature}. The activity-state detector is trained using Home windows as idle samples and target-application windows as active samples. 
Both inference stages are evaluated with three supervised classifiers: Random Forest (RF), XGBoost, and Support Vector Machine (SVM). These models provide complementary baselines for thermal application fingerprinting. RF and XGBoost capture non-linear feature interactions among spatial and temporal thermal features, while SVM provides a margin-based classifier for comparison.
Preprocessing parameters (standardization, low-variance filtering using a 40~mK threshold corresponding to the camera's noise-equivalent temperature difference, and ANOVA F-score feature selection) are fit on the training split and applied unchanged to held-out data. Full preprocessing details are deferred to \cref{app:preprocessing}.

\section{Experimental Setup}
\label{sec:experiments}

We evaluate \name across three experiments (E1--E3) corresponding to the three research questions of \cref{sec:rqs}.
\textbf{E1} establishes baseline indoor effectiveness and supports our feature-design ablations (grid resolution, window length, classifier choice). \textbf{E2} evaluates outdoor robustness and the indoor-to-outdoor domain shift. \textbf{E3} evaluates whether thermal signatures transfer across headset models or are dominated by device-specific characteristics.

\subsection{VR Devices and Applications}
\label{subsec:vr_apps}
Our primary evaluation device is a Meta Quest 3 \cite{meta_quest_3_2023}, a consumer-grade commercial-off-the-shelf (COTS) standalone VR headset. For cross-device experiments, we additionally evaluate two other VR headsets: the Meta Quest 2 \cite{meta_quest_2_2020}  and the HTC Vive Focus Vision\cite{htc_vive_focus_vision_2024}. The application test suite is selected to span idle, media, communication, productivity, and simulation workloads (see \cref{tab:app_suite_main}). Most applications are standalone application, except Zoom web for which we utilized the in-built browser to access it in every headset.

\begin{table}[htb]
    \centering
    \small
    \caption{Application test suite used in experiments.} 
    \label{tab:app_suite_main}
    \begin{tabular}{lll}
        \hline
        \textbf{Application} & \textbf{Category} & \textbf{Observed load} \\
        \hline
        Home & System & Baseline idle \\
        YouTube\cite{youtube_app} & Media & Low \\
        Media Player & Media & Low \\
        Zoom Web\cite{zoom_web} & Communication & Low-medium \\
        Arkio\cite{arkio_app} & Productivity & Medium \\
        First Hand\cite{firsthand_vr} & Simulation & High \\
        VRFS\cite{vrfs_app} & Simulation & High \\
        \hline
    \end{tabular}
\end{table}

\begin{figure}[htb]
    \centering
    
    \begin{subfigure}[t]{0.50\linewidth}
        \centering
        \includegraphics[width=\linewidth]{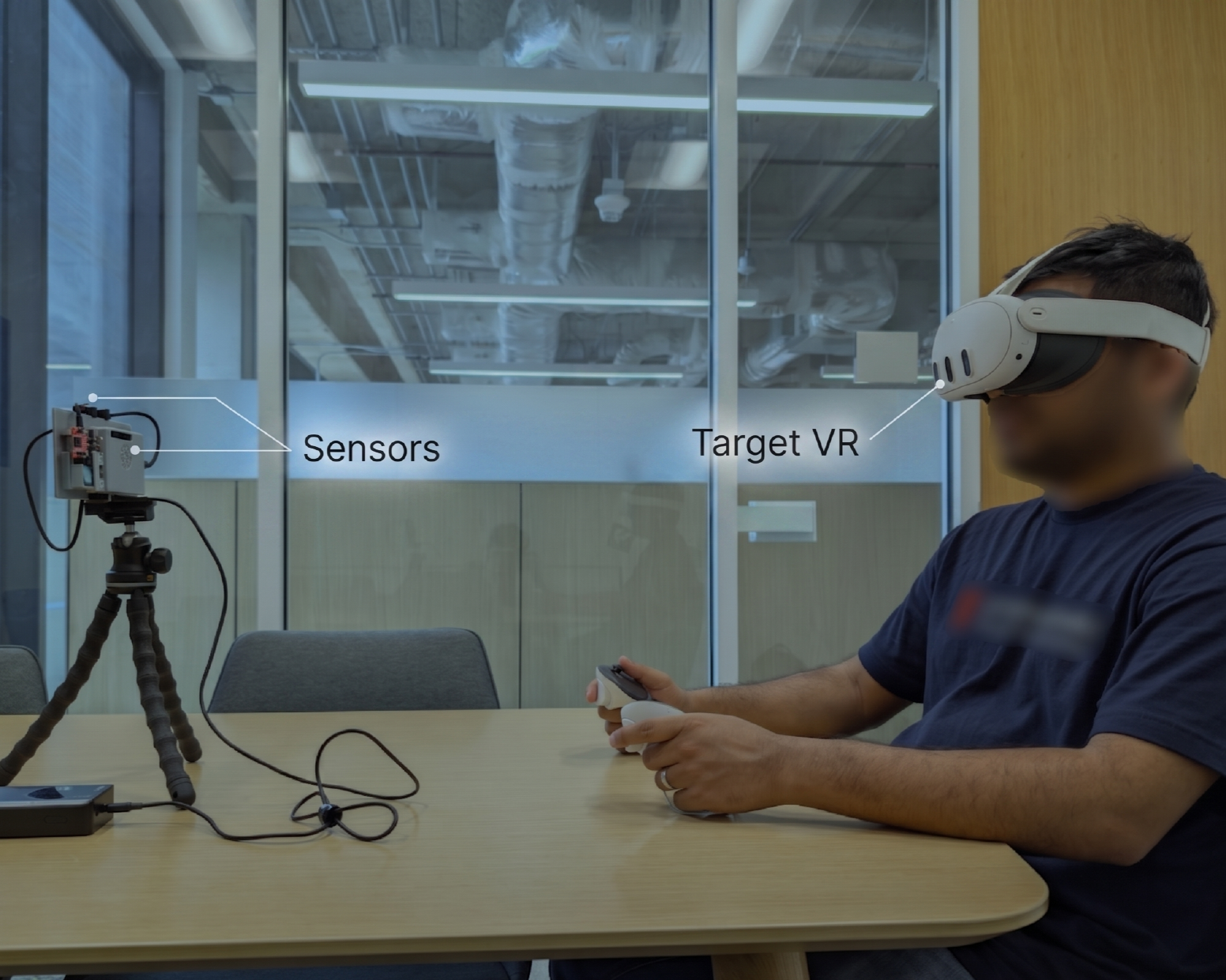}
        \caption{}
        \label{fig:data_collection}
    \end{subfigure}
    \hfill
    \begin{subfigure}[t]{0.465\linewidth}
        \centering
        \includegraphics[width=\linewidth]{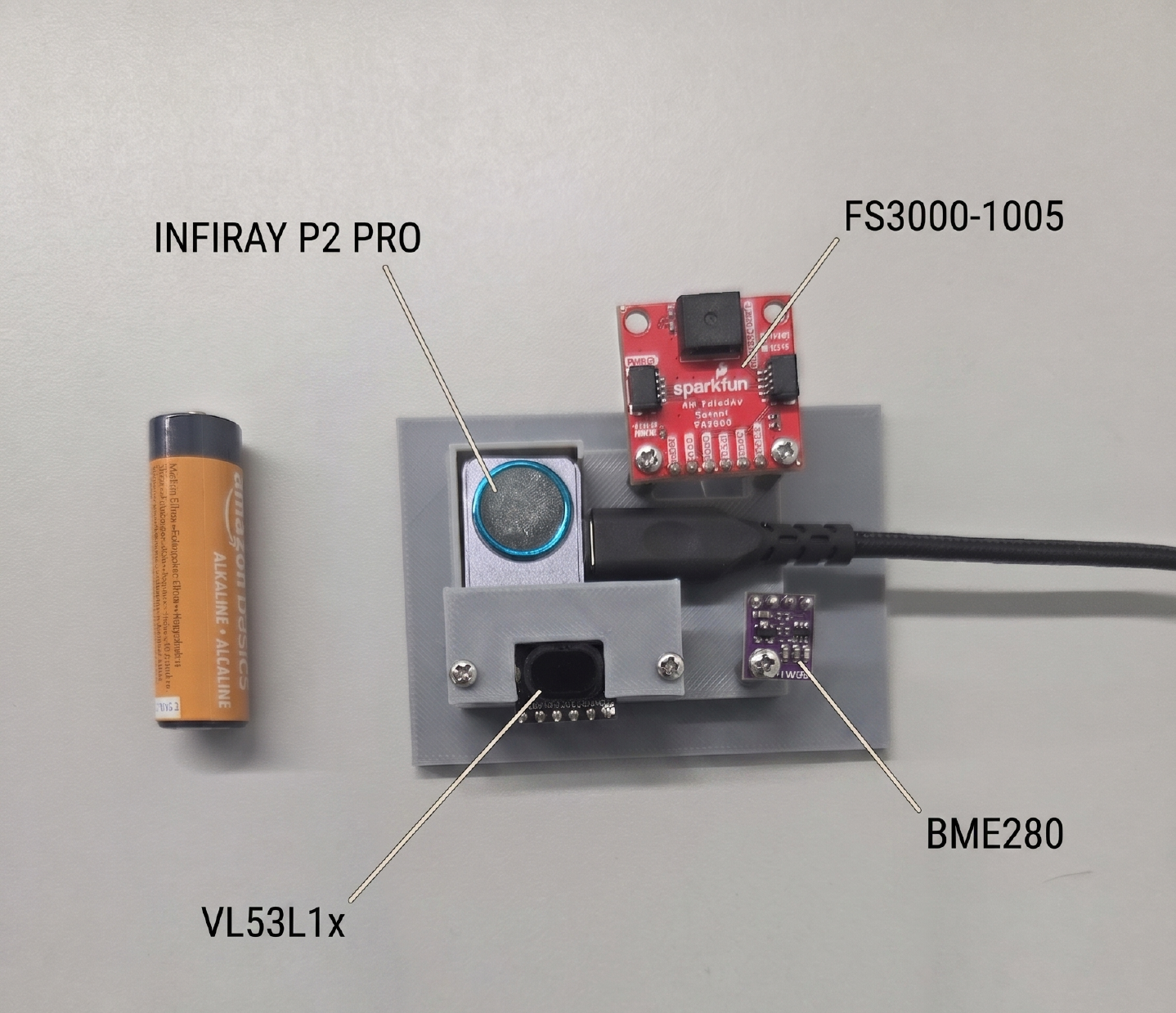}
        \caption{}
        \label{fig:pi_tool}
    \end{subfigure}
    
    \caption{(a) Indoor experimental setup, (b) data acquisition apparatus.}
    \label{fig:data_collection_setup}
\end{figure}

\subsection{Acquisition Apparatus}
\label{subsec:apparatus}

\name operates on calibrated \emph{radiometric} frames, where each pixel encodes an absolute temperature value rather than a colorized intensity. This requirement emerged from our initial experiments with USB Video Class (UVC) based capture on consumer thermal cameras such as the TopDon TC001: the UVC stream exposed a vendor-rendered RGB heatmap with auto-gain and color mapping, rather than stable per-pixel temperatures. Such visualization streams confound true device heat with rendering choices and cannot be reliably subtracted, normalized, or compared across recordings.
To capture radiometric data without depending on a vendor application, we built a Raspberry Pi based apparatus. It centers on an Infiray P2 Pro thermal camera~\cite{infiray_p2_pro_thermal_camera}, selected for its open SDK, and stores each frame as both a thermal image and a CSV matrix of per-pixel temperatures sampled at 1~Hz.
The apparatus also integrates three additional sensors that record context observed in \cref{subsec:pilot_observations} to alter VR device surface temperature independent of the executing application:

\begin{enumerate}[leftmargin=*]
    \item \textbf{BME280.} Records ambient temperature which affect heat dissipation and radiative contrast between the headset and background.
    \item \textbf{FS3000-1005.} Measures air velocity, allowing \name to quantify convective cooling caused by indoor airflow or outdoor wind.
    \item \textbf{VL53L1x.} Measures camera-to-headset distance using time-of-flight (ToF) sensing, providing a continuous geometric context variable for each captured frame.
\end{enumerate}

The sensing component of the apparatus is compact and low-profile; as shown in \cref{fig:pi_tool} (next to AA battery for scale). Since the Raspberry Pi can be connected by cable and placed separately, the visible sensing component can remain small and inconspicuous, enabling realistic deployment in shared observation settings.
All measurements from these sensors are collected and timestamped on the Raspberry Pi so each thermal frame is paired with the contemporaneous environmental reading. Because the apparatus uses only commodity components, the attacker requires no specialized laboratory equipment or physical instrumentation of the victim device.

\subsection{Data Collection Protocol}
\label{subsec:data_collection_protocol}

Data collection was conducted by members of the research team. Because \name measures heat emitted from the headset chassis rather than biometric, behavioral, or interaction-specific attributes of the wearer, we did not conduct a participant-centered user study. The wearer served only as an operator to reproduce a realistic headset-use setting, with the headset worn in a normal seated posture and facing the thermal camera. Thus, the collected signal is device- and application-centered rather than participant-dependent. Based on this study design and our university's institutional policies, IRB approval for human-subject data protection was not required.
For each recording session, the \name apparatus is placed in front of the headset at the target distance and records synchronized thermal and environmental measurements. Each session follows the same four-phase protocol:

\begin{enumerate}[leftmargin=*]
    \item \textbf{Baseline.} The headset remains idle at the home screen until the initial thermal state stabilizes, allowing us to observe the starting thermal state before launching the target application.
    \item \textbf{Heat-up.} The target application is launched and used while the headset surface temperature begins to rise in response to the new workload.
    \item \textbf{Steady state.} The application continues under typical use while \name records the main thermal signature used for classification.
    \item \textbf{Cool-down.} The application is exited and the headset returns toward its idle thermal state, allowing us to observe residual heat and thermal hysteresis.
\end{enumerate}

Thermal data is captured at a sampling rate of \SI{1}{\hertz}. Each session corresponds to a single recording trial for one target application (or the idle Home state) and lasts approximately 10 minutes, yielding about 600 thermal frames. Each frame is stored as a per-pixel temperature matrix and synchronized with environmental sensor readings (captured via the auxiliary sensors mentioned in \cref{subsec:apparatus}).

\subsection{Dataset Summary}
\label{subsec:dataset_summary}

\Cref{tab:dataset_summary} summarizes the three datasets used in our evaluation. For experiment E1, we collected 67 indoor sessions. The per-class distribution is Home (11), YouTube (11), Media Player (11), Zoom Web (8), Arkio (8), First Hand (8), and VRFS (10). For E2, we collected 28 outdoor sessions containing four sessions per class across the same seven applications used in E1. For E3, we collected 84 indoor sessions across. The per-class distribution is Home (12), YouTube (12), Media Player (12), Zoom Web (12), Arkio (8), First Hand (8), VRFS (8), and Open Brush (12). The per-device distribution is Meta Quest 3 (32), Meta Quest 2 (32), and HTC Vive Focus Vision (20). 
E3 includes Open Brush as a replacement app since HTC ecosystem does not have the same set of apps as Meta.

\begin{table}[htb]
    \centering
    \small
    \caption{Dataset summary.}
    \label{tab:dataset_summary}
    \begin{tabularx}{\columnwidth}{llcX}
        \toprule
        \textbf{Dataset} & \textbf{Environment} & \textbf{Sessions} & \textbf{Purpose} \\
        \midrule
        E1 & Indoor & 67 & Main fingerprinting evaluation and ablations. \\
        E2 & Outdoor & 28 & Environmental robustness under outdoor conditions. \\
        E3 & Multiple devices & 84 & Cross-device transfer and device-level fingerprinting. \\
        \bottomrule
    \end{tabularx}
\end{table}

\subsection{Evaluation Protocol}
\label{subsec:evaluation_protocol}
For E1, we use leave-one-session-out (LOSO) cross-validation. In each fold, one complete recording session  is held out for testing, while the remaining sessions are used for training. All preprocessing steps are fit only on the training sessions in that fold and then applied to the held-out session. This prevents leakage from temporally adjacent or overlapping windows from the same session.

For E2, we evaluate indoor-to-outdoor robustness under two settings. In the \emph{zero-shot transfer} setting, the model is trained only on controlled indoor sessions from E1 and tested directly on outdoor sessions from E2. This measures whether thermal signatures learned indoors transfer to outdoor conditions without any outdoor training examples. In the \emph{few-shot outdoor adaptation} setting, a limited number of outdoor sessions are added to the indoor training set, while disjoint outdoor sessions are reserved for testing. This measures whether small amounts of target-environment data reduce the indoor-to-outdoor domain shift.

For E3, we evaluate cross-device behavior under two settings. In the \emph{pooled cross-device} setting, sessions from all headset models are included in the dataset and evaluated using LOSO cross-validation. This measures fingerprinting performance when the model has seen training examples from each headset family. In the \emph{leave-one-device-out} (LODO) setting, all sessions from one headset model are held out for testing, while sessions from the remaining headset models are used for training. This measures whether application signatures transfer to an unseen headset model.

\subsection{Metrics}
\label{subsec:metrics}

We report accuracy, precision, recall, and weighted F1-score for application fingerprinting experiments. We use weighted F1-score as the primary summary metric because it balances precision and recall while accounting for class support.
To quantitatively assess the segmentation models, we utilized a combination of overlap and distance metrics: the Dice Coefficient (DSC) measures the global degree of spatial overlap between the prediction and ground truth (bounded 0 to 1); the Jaccard Index (IoU) provides a complementary measure of overlap severity (bounded 0 to 1); the 95\% Hausdorff Distance (95HD) calculates the 95th percentile of the shortest distances between all predicted and true surface boundary points to evaluate contour precision while resisting outliers (lower is better); and the Average Surface Distance (ASD) computes the mean of the shortest distances between the predicted and true surface boundaries to indicate how tightly the contours match on average (lower is better).

\section{Evaluation}
\label{sec:results}

We organize the evaluation around the three experiments E1--E3 introduced in \cref{sec:experiments}. Before reporting the fingerprinting results, we briefly establish that the upstream ROI segmentation step (\cref{subsec:roi_segmentation}) is reliable enough to support the downstream analysis. We use experiment E1 to compare window sizes, grid resolutions and classifier choices. The best-performing configuration from E1 is used for the subsequent E2 and E3 experiments.

\subsection{ROI Segmentation Performance}
\label{subsec:seg_results}

\begin{figure}[htb]
    \centering
    \includegraphics[width=1\linewidth]{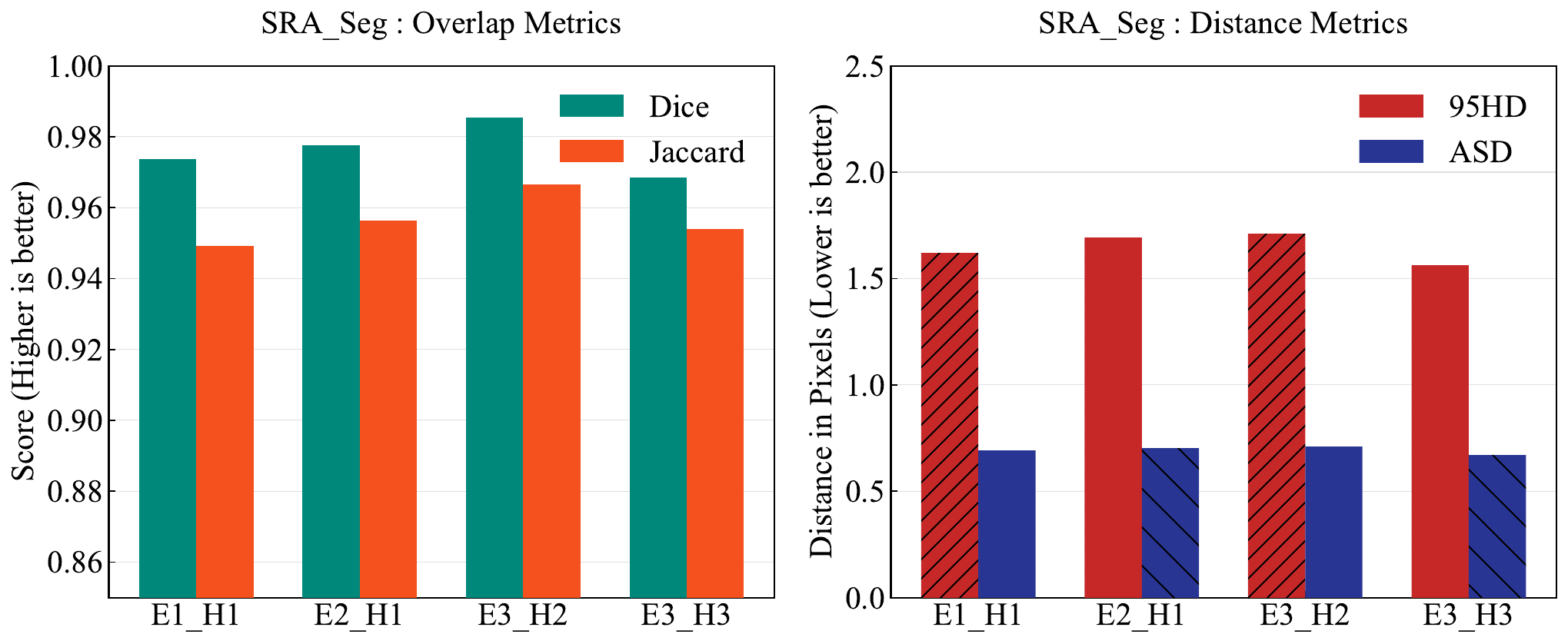}
    \caption{Segmentation model performance.}
    \label{fig:seg_sra_eval}
\end{figure}

We evaluate the SRA-Seg-based ROI segmentation pipeline described in \cref{subsec:roi_segmentation}. Because the three headset models differ in shape, scale, and thermal appearance, we train separate segmentation models for H1 (Meta Quest 3), H2 (Meta Quest 2), and H3 (HTC Vive Focus Vision). The initial H1 model, trained on indoor E1 data, generalized well within E1 but did not reliably isolate the headset under the outdoor conditions in E2. We therefore retrained the H1 model with additional manually annotated E2 frames and use this updated model for both E2 and the H1 sessions in E3. Separate models are trained for H2 and H3.
As shown in \cref{fig:seg_sra_eval}, all models achieve high overlap with the manually annotated masks. The H2 model obtains the highest overlap scores, with a Dice coefficient of 98.55\% and a Jaccard index of 96.65\%. Although H2 also has slightly higher 95HD and ASD, the differences are small and do not materially affect downstream feature extraction. Overall, these results indicate that the ROI segmentation stage reliably isolates the headset surface across devices and environments.

\subsection{E1: Indoor Experiments}
\label{subsec:e1_results}

\begin{figure}
    \centering
    \includegraphics[width=0.99\linewidth]{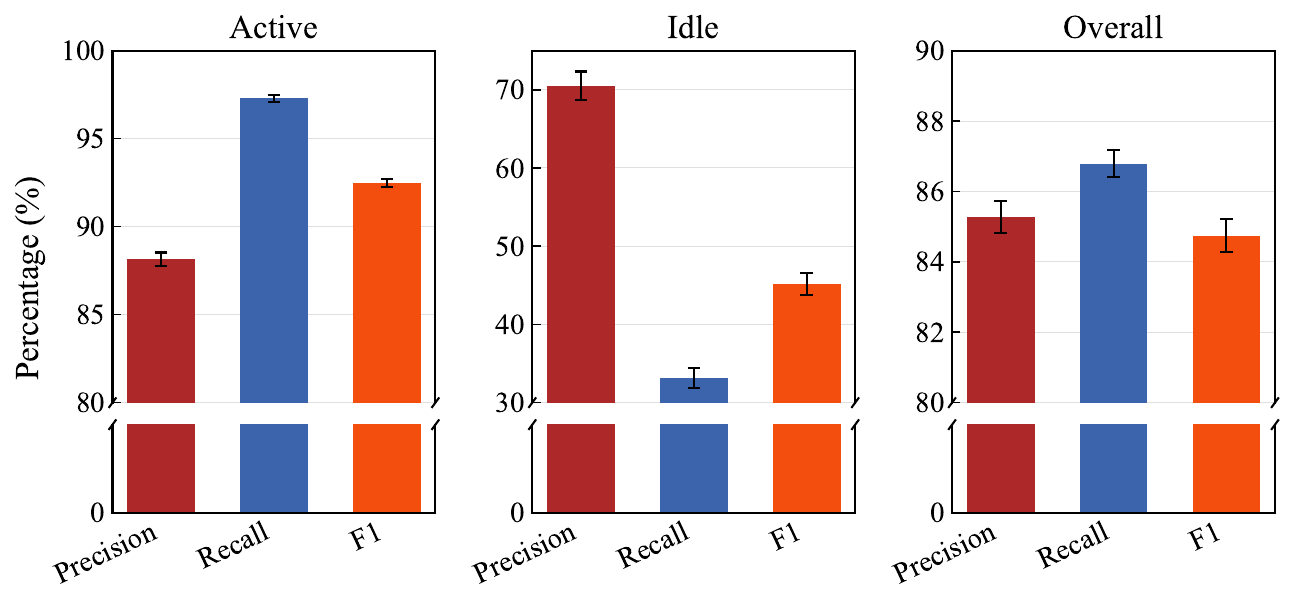}
    \caption{Activity-state detection.}
    \label{fig:activity_state_barplot}
\end{figure}

\noindent\textbf{Stage 1: Activity-state Detection.}
We first evaluate the first stage of \name's two-stage inference pipeline: distinguishing idle headset behavior from active application use. 
As shown in \cref{tab:gk_perf,fig:activity_state_barplot}, the Random Forest detector achieves high recall for active use (97.3\%, $\sigma=0.19$), which is the critical operating point for the attack because active windows must reach the application-fingerprinting stage. Idle/Home windows are harder to identify, with 33.2\% recall ($\sigma=1.29$), reflecting the fact that Home is not a stable workload and can be affected by residual heat and background system activity. In this pipeline, an idle window misclassified as active only causes an unnecessary Stage~2 prediction, whereas an active window misclassified as idle prevents fingerprinting altogether. We therefore prioritize high active-use recall over balanced idle/active performance.

\begin{figure}[htb]
    \centering
    \includegraphics[width=0.99\linewidth]{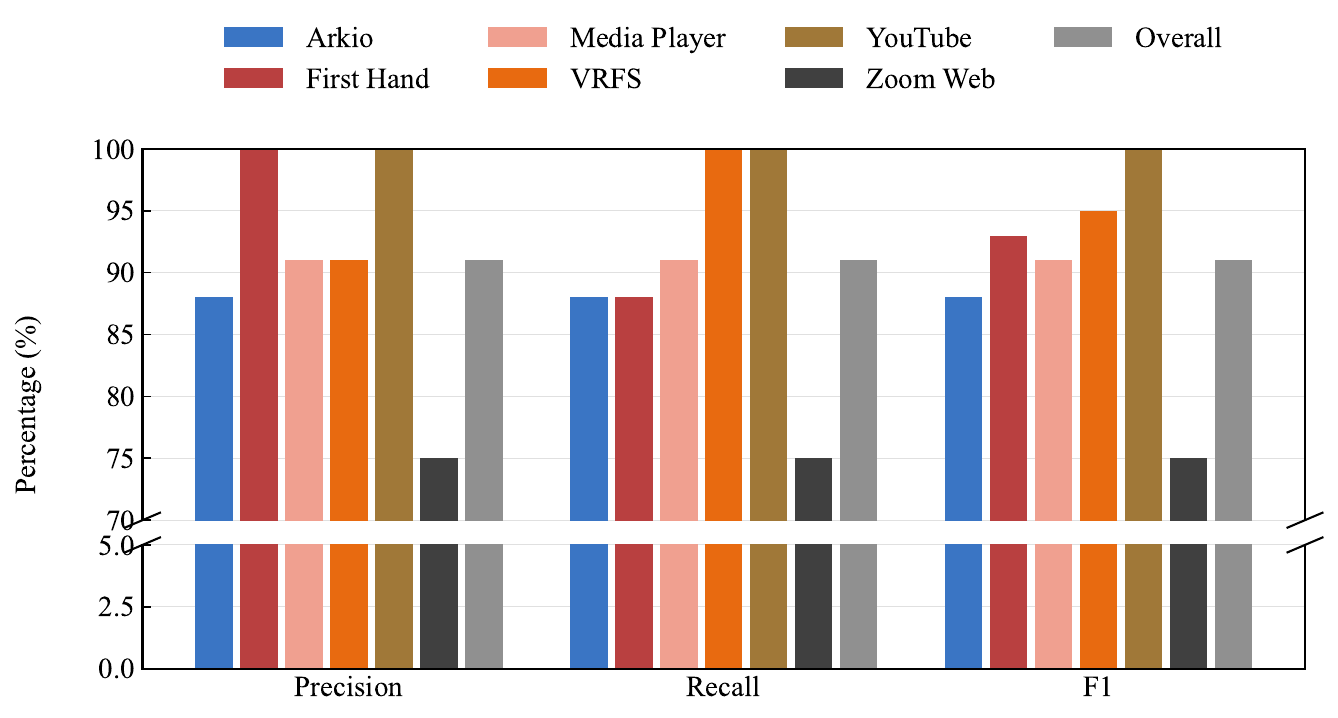}
    \caption{Per-app performance of \name in E1.}
    \label{fig:per_app_comp}
\end{figure}

\smallskip
\noindent\textbf{Stage 2: Active-app Fingerprinting.} 
For stage 2. we use a $16\times16$ grid representation, \SI{10}{s} prediction windows, and a Random Forest classifier. A detailed analysis of these configurations are provided in \cref{subsec:ablations}.
\name achieves strong window-level performance across the six active VR applications (see \cref{fig:per_app_comp}), with mean per-app F1-score of 91\%.

\begin{figure}[htb]
    \centering
    \includegraphics[width=0.7\linewidth]{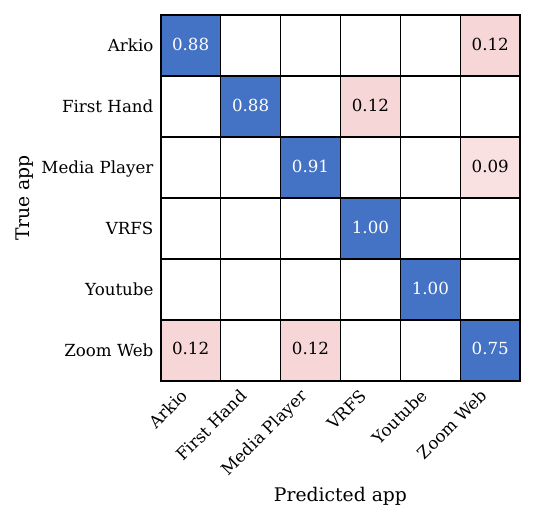}
    \caption{Confusion matrix for indoor experiment E1.}
    \label{fig:conf_e1}
\end{figure}

The strongest performance occurs for YouTube, VRFS, and First Hand. YouTube is classified perfectly across all three metrics, while VRFS and First Hand remain highly distinguishable, achieving F1-scores of 95\% and 93\%, respectively.
These results suggest that applications whose workloads remain relatively stable within a window are easier to identify from thermal emissions. For example, YouTube playback and simulation workloads such as VRFS and First Hand continuously exercise the display, rendering pipeline, and processor, producing repeatable heating patterns over short windows.
In contrast, Zoom Web is the weakest class (F1=75\%). One likely explanation is that its workload is less stable over shorter windows than continuous playback or simulation workloads, causing its thermal signature to overlap with other applications. The confusion matrix in \cref{fig:conf_e1} shows that Zoom Web is most often confused with the other low-medium load classes (Arkio and Media Player), consistent with this interpretation.

We also report our classifier performance that includes Home alongside the six active applications (without running idle-detection step). \Cref{tab:per_class_7class} shows that the Home state is substantially harder to classify, with an F1-score of only 17\% with mean per-app F1-score dropping to 76\%. This is expected because Home is not a stable/unique application workload; it is an idle/system state whose thermal profile depends strongly on residual heat, cooling trajectory, and the preceding session. This shows the effectiveness of our the hierarchical formulation in which activity-state detection is separated from active-application fingerprinting.

\subsection{Configuration Ablations}
\label{subsec:ablations}

\begin{figure}[htb]
    \centering
    \includegraphics[width=0.7\linewidth]{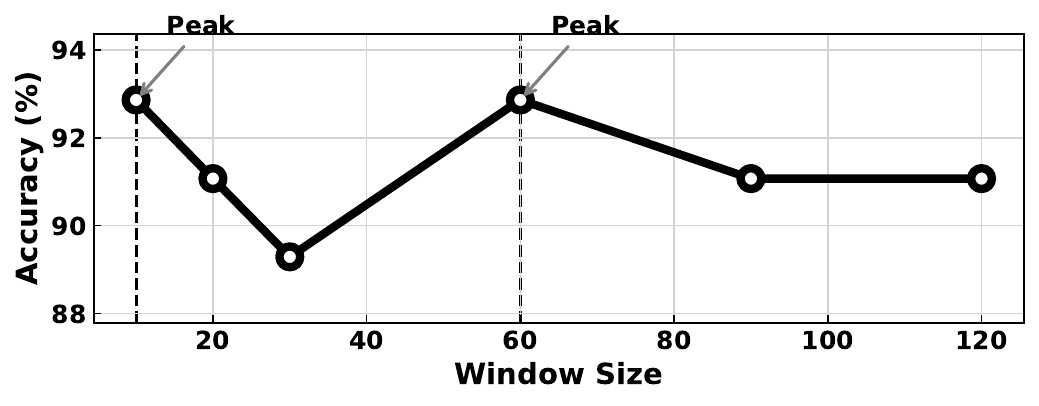}
    \caption{Window size vs. accuracy.}
    \label{fig:window_size}
\end{figure}

\noindent\textbf{Effect of Window Size.}
We further evaluate active-application fingerprinting using fixed-length time windows of 10, 20, 30, 60, 90, and 120 seconds, i.e., the observation duration required by an attacker to infer a running application of a victim . Across these window sizes, \name achieves comparable mean per-app accuracy of approx. 90\% across all apps (see \cref{fig:window_size}). 
This suggests that short windows already capture sufficient application-induced thermal structure, and longer windows provide limited additional benefit. 
We therefore use \SI{10}{s} windows for all our subsequent experiments.

\smallskip
\noindent\textbf{Effect of Grid Resolution.}
We then evaluate the grid resolution parameter introduced in \cref{subsec:thermal_signature}.
We sweep the grid sizes from \(4 \times 4\) to \(24 \times 24\), using the Random Forest classifier described in \cref{sec:classification}.

\begin{figure}[htb]
    \centering
    \includegraphics[width=0.7\linewidth]{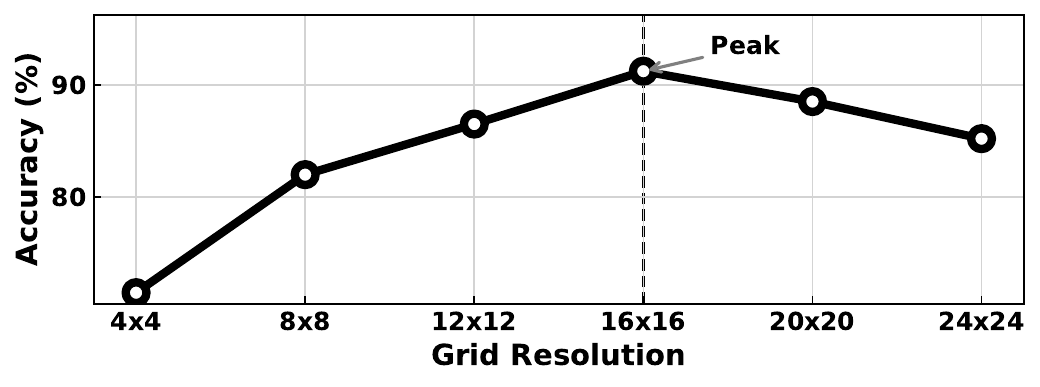}
    \caption{Grid resolution vs. accuracy.}
    \label{fig:grid_resolution}
\end{figure} 

As shown in \cref{fig:grid_resolution}, performance improves as the representation preserves more localized thermal structure, reaching its best accuracy at \(16 \times 16\) with a mean per-app accuracy of 91.23\%. Beyond this point, performance declines, suggesting that very fine grids introduce noise and session-specific artifacts that outweigh the benefit of additional spatial detail. We therefore use the \(16 \times 16\) grid as the default representation in subsequent experiments.

\smallskip
\noindent\textbf{ML Model Comparison}
We then compare three supervised classifiers: Random Forest, XGBoost, and SVM (using a \(16 \times 16\) grid representation and \SI{10}{s} window size). 

\begin{figure}[htb]
    \centering
    \includegraphics[width=0.8\linewidth]{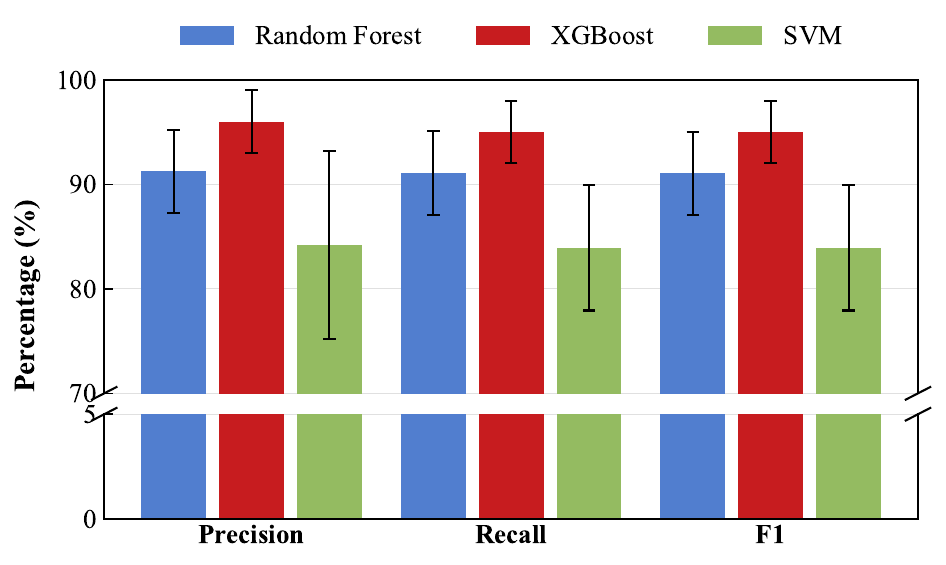}
    \caption{ML model performance comparison.}
    \label{fig:model_comp}
\end{figure}
\begin{figure}[htbp]
    \centering
    \includegraphics[width=0.85\linewidth]{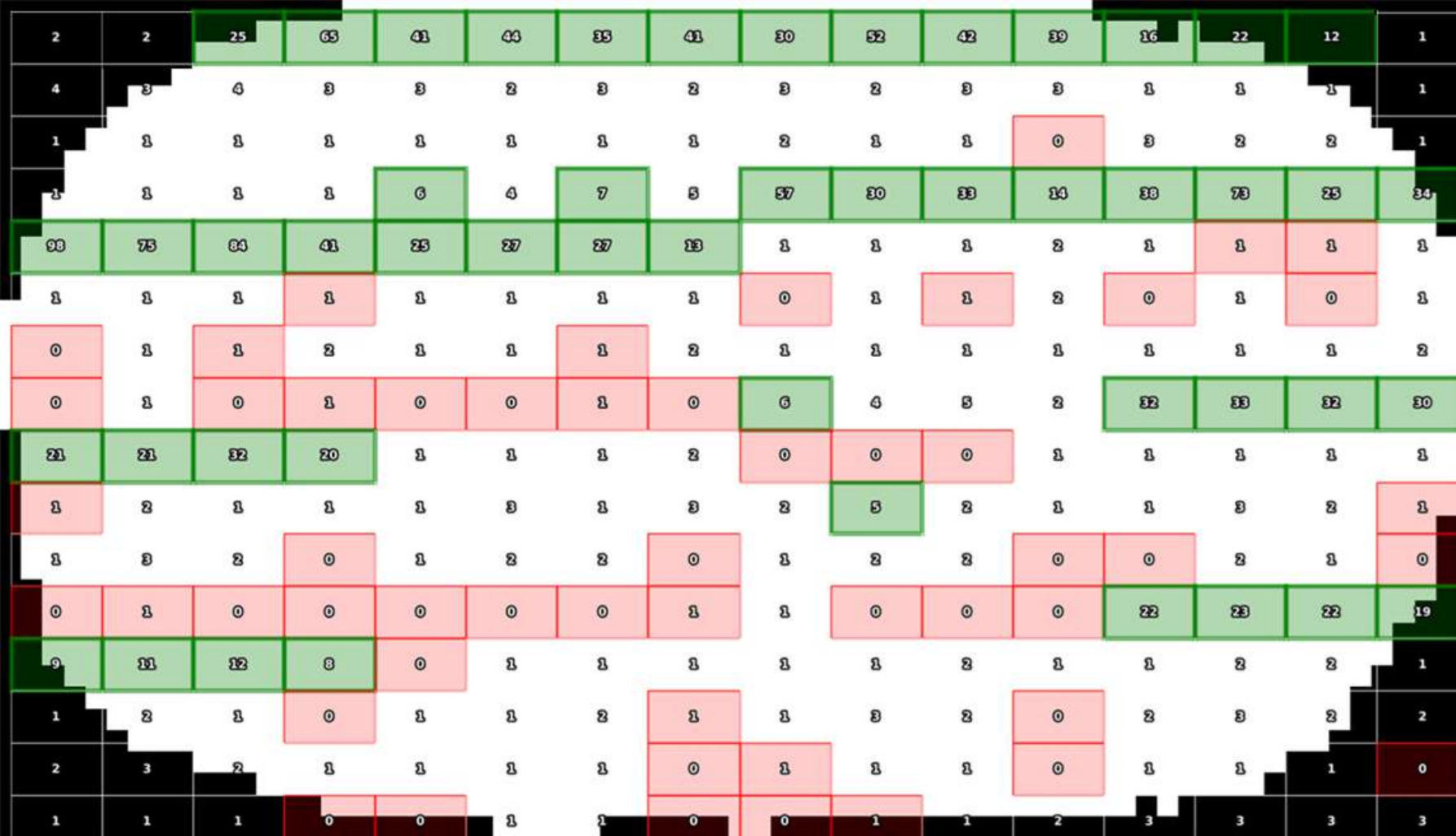}
    \caption{16x16 Grid's mask importance region.}
    \label{fig:mask_region_importance}
\end{figure}

 
\Cref{fig:model_comp} shows that XGBoost achieves the highest mean per-app F1-score (95\%), Random Forest is close behind (91\%), and SVM trails (84\%) with markedly higher variance across folds ($\sigma=6-9$). These results confirm that application-specific information is reliably present in the thermal signatures. Tree-based ensembles are highly effective for \name's heterogeneous statistical, spatial, and temporal features, with XGBoost achieving the highest overall accuracy. However, we intentionally utilize Random Forest as our primary classifier for the remainder of the experiments to prioritize model stability and spatial explainability over marginal performance gains. Unlike boosting techniques, which are highly sensitive to noisy data and outliers \cite{dietterich2000experimental}, Random Forest's bagging architecture averages across uncorrelated decision trees. This makes it inherently resistant to the environmental noise \cite{breiman2001random}. 
Furthermore, it naturally yields the stable feature importance mappings required for our spatial leakage analysis (\cref{fig:mask_region_importance}), all while comfortably exceeding the attack viability threshold of 90\% accuracy.

\begin{figure}[h]
    \centering
    \includegraphics[width=0.53\linewidth]{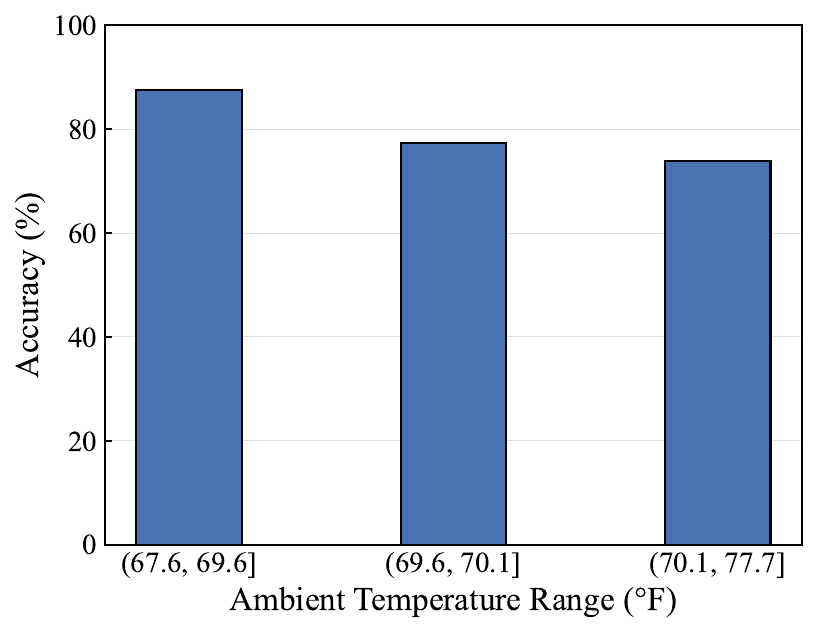}
    \caption{Accuracy vs temperature.}
    \label{fig:temp_acc}
\end{figure}

\smallskip
\noindent\textbf{Effect of Environmental Factors.}
We also examine how environmental variation affects app classification performance. Across the E1 dataset, the highest mean per-app accuracies occur when ambient temperature and camera distance fall within narrower ranges: 67.6°F to 69.6°F (19.8°C to 20.9°C) for ambient temperature and \SI{45.8}{cm} to \SI{60.2}{cm} for camera distance. Within these ranges, accuracy improves by approximately 5--8\% relative to boundary conditions (see \cref{fig:temp_acc,fig:dist_acc}). However, even at an extended observation distance of \SI{100}{cm}, the model maintained a robust application fingerprinting accuracy of approximately 80\%.
These results show that even under indoor conditions, thermal fingerprinting is affected by measurement geometry and ambient temperature. This motivates the environmental sensing components of \name (detailed in \cref{subsec:env_normalization}) and directly informs E2, where we evaluate whether these effects can be compensated under highly variable outdoor conditions.

\smallskip
\noindent\textbf{Feature Importance.}
Finally, we analyze which spatial regions contribute most to classification. For each LOSO fold, we extract Random Forest feature importance scores, which quantify the relative contribution of each grid cell to the classifier's decisions, and map them back to the \(16 \times 16\) headset grid.
The importance heatmap \cref{fig:mask_region_importance} shows elevated importance in cells (green highlighted ones are in top 20\%, red ones are in bottom 20\%) corresponding to:

\begin{enumerate}[leftmargin=*]
    \item \textbf{Central Processing Region:} Cells overlaying the approximate SoC location exhibit high importance, reflecting the direct relationship between computational load and localized heating.
    \item \textbf{Exhaust Vent Regions:} Cells near the top of the headset (where exhaust vents are typically located) show distinctive patterns, as fan-driven convective cooling creates characteristic thermal gradients that vary with application intensity.
    \item \textbf{Battery Region:} Moderate importance is observed in cells corresponding to the battery location, as power draw correlates with application demands.
\end{enumerate}
This demonstrate that \name is learning physically meaningful thermal patterns rather than arbitrary image artifacts. In particular, applications with higher computational demand produce localized heating and cooling dynamics that are captured by the spatiotemporal grid representation.

\noindent\textbf{\emph{E1 Summary:}} Externally emitted headset heat carries application-specific information sufficient to fingerprint six active VR applications with above 90\% accuracy under indoor conditions, with the discriminative signal concentrated on physically interpretable regions of the chassis.

\subsection{E2: Outdoor Robustness and Environmental Variability.}
\label{subsec:e2_results}

E2 evaluates the robustness of \name in outdoor environments. We use the $16\times16$ grid size and Random Forest model configuration to evaluate three settings: zero-shot indoor-to-outdoor transfer, few-shot outdoor adaptation, and few-shot outdoor adaptation with environmental normalization (\cref{subsec:env_normalization}).

\begin{figure}
    \centering
    \includegraphics[width=0.8\linewidth]{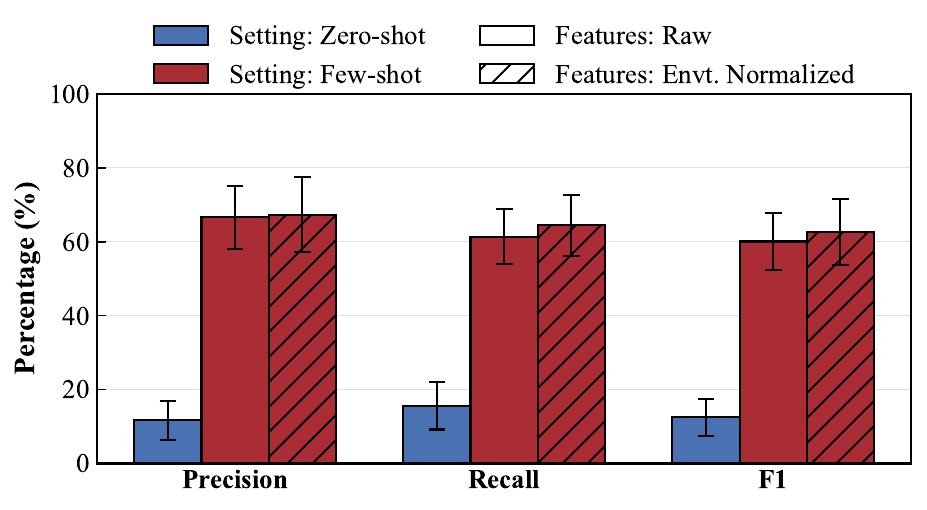}
    \caption{Effect of normalization on outdoor experiments.}
    \label{fig:outdoor_norm}
\end{figure}

\Cref{fig:outdoor_norm} summarizes the outdoor app-classification results. Direct indoor-to-outdoor transfer performs poorly: when trained only on indoor E1 sessions and tested on outdoor E2 sessions, \name achieves only 12.49\% mean F1 ($\sigma=4.95$). This confirms a substantial domain shift between controlled indoor and outdoor thermal measurements, where ambient temperature variation, natural airflow, and background heating alter the observed thermal patterns independently of the running application.
However, adding a small number of outdoor sessions to the training set considerably improves application classification, increasing mean F1 to 60.06\% ($\sigma=7.78$). Further, applying environmental normalization provides a smaller but consistent additional improvement, increasing mean F1 to 62.55\% ($\sigma=8.88$). Thus, few-shot outdoor calibration accounts for most of the recovery from domain shift, while environmental normalization provides additional robustness.

\begin{figure}[htb]
    \centering
    \includegraphics[width=0.9\linewidth]{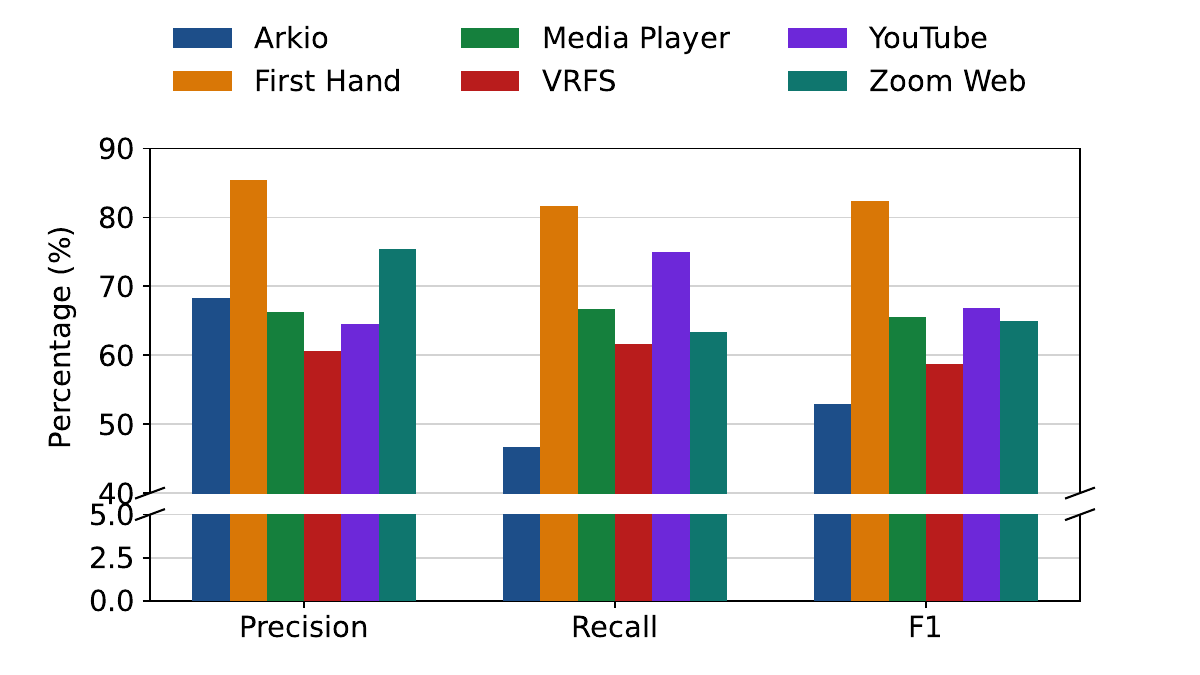}
    \caption{E2 session level performance.}
    \label{fig:e2_session_per_app}
\end{figure}

We further evaluate E2 using session-level aggregation. Each outdoor recording session contains multiple prediction windows; \name classifies each window separately and then assigns the session to the application predicted by the majority of its windows.
This analysis complements the window-level results: the window-level setting captures the minimum observation granularity available to the attacker, while session-level aggregation reflects an attacker who can observe the same application continuously over a longer interval. Session-level aggregation is especially relevant outdoors because short-term perturbations, such as airflow changes, transient background heating, and occasional segmentation noise, can corrupt individual windows without necessarily changing the dominant thermal signature over the full application session.

As shown in \cref{tab:e2_session_per_app}, session-level aggregation improves the stability of outdoor inference, reaching a mean F1-score of 65.25\% across the six active applications. These results suggests that some outdoor errors are transient at the window level and can be reduced when predictions are aggregated over a full recording session.
The per-app results also show that outdoor degradation is application-dependent. First Hand is the strongest outdoor case, with an F1-score of 82.32\%. 
Several other applications also retain useful signal in at least one direction: Zoom Web achieves 75.42\% precision, YouTube achieves 75.00\% recall, and Arkio achieves 68.33\% precision. Overall, \name retains useful outdoor signal at the session level, but environmental variability still reduces class separability compared with the indoor setting. 

\noindent\textbf{\emph{E2 Summary.}} Outdoor conditions introduce substantial domain shift, reducing application-classification performance under direct indoor-to-outdoor transfer. Few-shot outdoor calibration recovers much of the lost signal, and environmental normalization provides additional robustness. Session-level inference further improves stability by suppressing transient outdoor errors.

\subsection{E3: Cross-Device Effects}
\label{subsec:e3_results}

E3 evaluates how headset-specific hardware characteristics affect thermal application fingerprinting. We conduct experiments using three VR headsets: Meta Quest 3, Meta Quest 2, and HTC Vive Focus Vision.
We evaluate two settings. In the pooled setting, training data includes sessions from all headset models. This corresponds to a device-aware attacker who can visually identify the victim headset model and has profiled the same headset family in advance. In the zero-shot setting, we use leave-one-device-out (LODO) evaluation, where all sessions from one headset model are held out for testing and the model is trained only on the remaining headset models.

\noindent\textbf{Pooled Cross-Device Fingerprinting.}
As shown in \cref{tab:e3_combined_results}, \name achieves strong pooled cross-device performance, reaching an F1-score of 81.08\% ($\sigma=0.19$). This result shows that application-level thermal information remains recoverable even when multiple headset types are present, provided that the model has seen representative data from each headset family during training.

\noindent\textbf{Zero-Shot Cross-Device Transfer.}
In the stricter LODO setting, app-inference performance drops sharply to only 11.59\% F1 ($\sigma=5.54$). 
The contrast between pooled and zero-shot performance reveals that externally observed thermal signatures are strongly influenced by device-specific factors, including hardware layout, cooling design, and chassis geometry. While application workload contributes to the thermal signal, its manifestation on the device surface depends on how heat propagates through the specific hardware.

\noindent\textbf{Headset-baseline Normalization.}
To test whether static headset-specific heat patterns explain the poor cross-device transfer, we also evaluate a diagnostic headset-baseline normalization. For each headset, we subtract an idle Home-state thermal baseline from active-application observations, producing residual features relative to that headset's resting thermal profile. The full procedure is described in \cref{app:headset_norm}.
As shown in \cref{tab:e3_combined_results}, headset-baseline normalization reduces pooled performance from a F1-score of 81.08\% to 34.68\%.
This suggests that pooled models benefit from device-conditioned application signatures, and that subtracting a static idle baseline is insufficient to make application signatures transfer across unseen headset models. In other words, device-specific thermal behavior is not limited to a fixed offset; it also affects how application-induced heat propagates during use.

\begin{table}[htb]
\centering
\small
\caption{Pooled vs. zero-shot results.}
\label{tab:e3_combined_results}

\begin{tabularx}{\columnwidth}{Xccc}
    \toprule
    \textbf{Setting} & \textbf{Precision \%} & \textbf{Recall \%} & \textbf{F1 \%} \\
    \midrule

    Pooled
    & 81.08 $\pm$ 9.17
    & 81.08 $\pm$ 9.17
    & 81.08 $\pm$ 9.17 \\

    Zero-shot
    & 11.39 $\pm$ 5.54
    & 14.29 $\pm$ 7.22
    & 11.59 $\pm$ 5.68 \\

    Pooled (Normalized)
    & 34.97 $\pm$ 7.74
    & 35.14 $\pm$ 7.74
    & 34.68 $\pm$ 7.74 \\

    \bottomrule
\end{tabularx}
\end{table}

\begin{figure}[htb]
    \centering
    \includegraphics[width=0.6\linewidth]{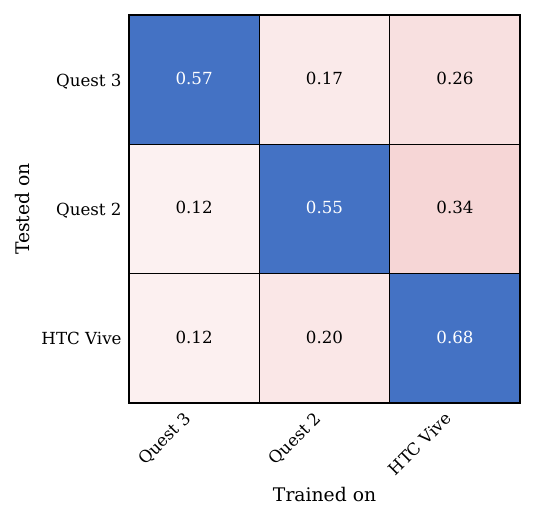}
    \caption{E3: Train-on-one-headset, test-on-another confusion matrix.}
    \label{fig:e3_conf_matrix}
\end{figure}

\noindent\textbf{Train-on-one-headset, Test-on-another.}
To further examine device dependence, we evaluate a stricter train-on-one-headset, test-on-another setting. 
As shown in \cref{fig:e3_conf_matrix}, same-headset performance is substantially higher than cross-headset transfer, but remains below the pooled setting. Off-diagonal results are consistently low, with the strongest transfer occurring from Meta Quest 2 to HTC Vive Focus Vision. These results reinforce the LODO finding: thermal signatures do not transfer reliably across headset models. They also suggest that pooling data from multiple headset families helps the classifier learn device-conditioned application signatures, whereas training on a single headset provides limited generalization.

\noindent\textbf{\emph{E3 Summary:}} Cross-device results show that thermal leakage exists across multiple commercial headsets, but accurate fingerprinting requires headset-model-specific training. These results indicate that \name is effective when the attacker has access to training data from the target headset model or similar devices. However, cross-device generalization without such data remains challenging, as hardware-specific thermal behavior dominates the externally observable signal.

\subsection{Factors Affecting Accuracy}
The results across E1--E3 show that \name's app-inference accuracy depends not only on whether applications generate distinct heat, but also on other external factors including observation stability and environmental factors.

\noindent\textbf{Head Pose and ROI Stability.}
In real-world VR headset usage, natural head movement changes the headset's position, scale, orientation, and visible boundary in the thermal frame. Small pose changes can therefore introduce feature variation unrelated to the running application, while larger rotations can hide thermally informative regions such as the SoC area, vents, or battery region. This effect is especially important for interactive VR appls, where users may naturally turn their head more frequently than during passive media playback.

\noindent\textbf{Partial Obstructions of the Headset.}
Because \name relies on line-of-sight thermal observation, temporary obstructions can degrade the measured signature. Actions such as hand movements, controller movement may partially block the front surface of the device. For example, simulation or design applications may induce more hand and head movement than video playback. As a result, obstructions can both remove useful thermal regions and introduce app-correlated artifacts that the classifier may incorrectly learn.

\noindent\textbf{Outdoor Environmental Effects.}
As discussed earlier, outdoor settings introduce temperature variation that is not caused by the application. Ambient temperature changes the contrast between the headset and background, airflow changes convective cooling, and solar exposure can create localized gradients unrelated to computation. Solar heating is especially difficult because it can heat one side of the headset more than the other depending on sun angle, reflections, and shadowing. Such patterns can resemble application-induced hotspots and therefore confuse the spatial grid features.

\subsection{Countermeasures}
We consider potential defenses informed by our findings.

\begin{figure}[htb]
    \centering
    \includegraphics[width=0.75\linewidth]{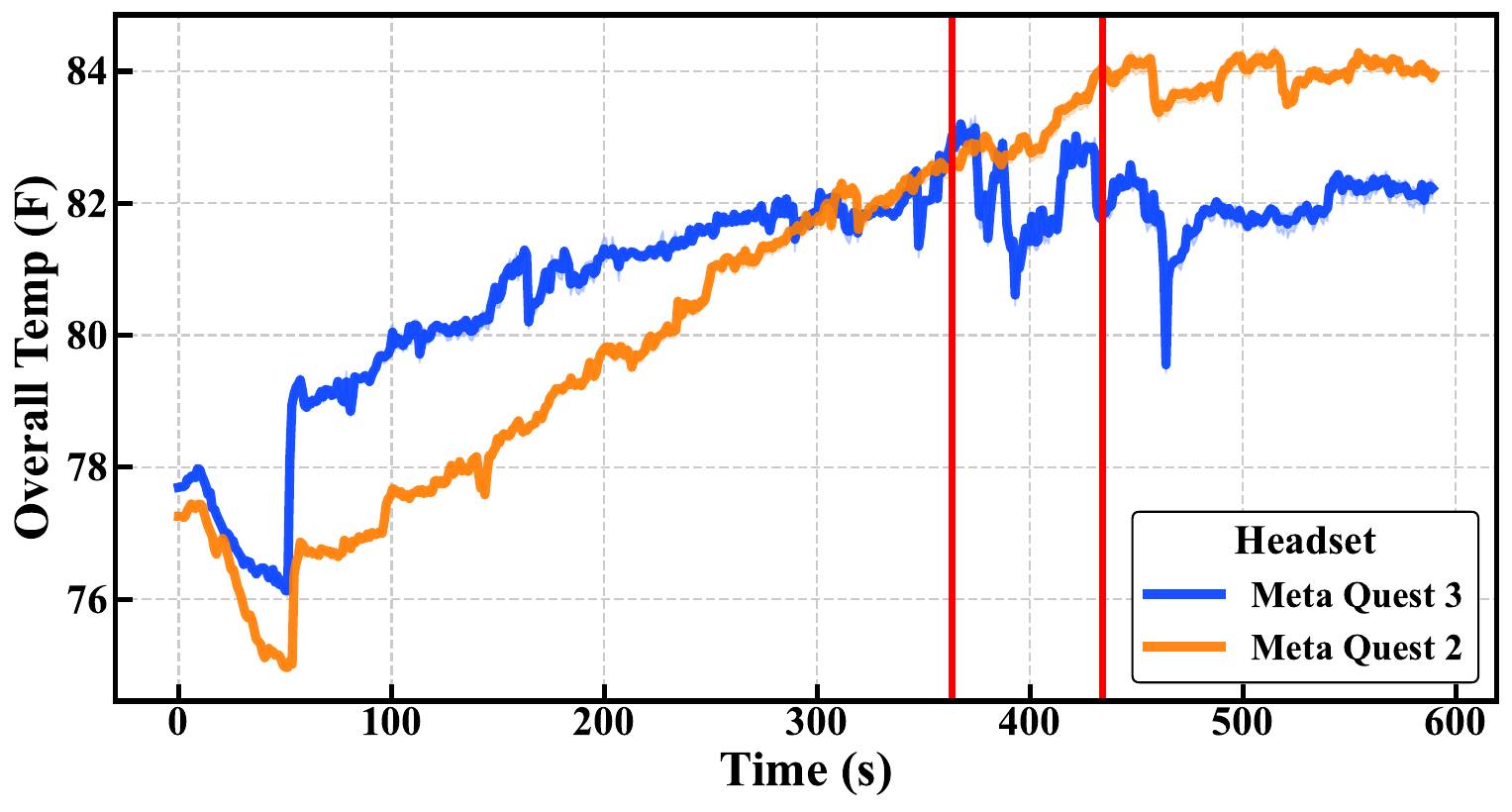}
    \caption{Effect of active cooling in an app session.}
    \label{fig:active_cool_temp_time}
\end{figure}

\noindent\textbf{Active Cooling.} Active cooling substantially obfuscates application thermal signatures in heavy-load regimes. \Cref{fig:active_cool_temp_time} contrasts a Meta Quest 2 session (which has only minimal active cooling) with a Meta Quest 3 session (which has a more aggressive fan profile). Once Quest~3's fan engages, the surface temperature trace shows characteristic dips that disrupt the otherwise-monotonic heat-up signal that downstream features rely on. Quest 2 fan engages later in the session at higher temperature, which creates significant difference in the overall thermal trend afterwards.

\begin{figure}[htb]
    \centering
    \includegraphics[width=0.75\linewidth]{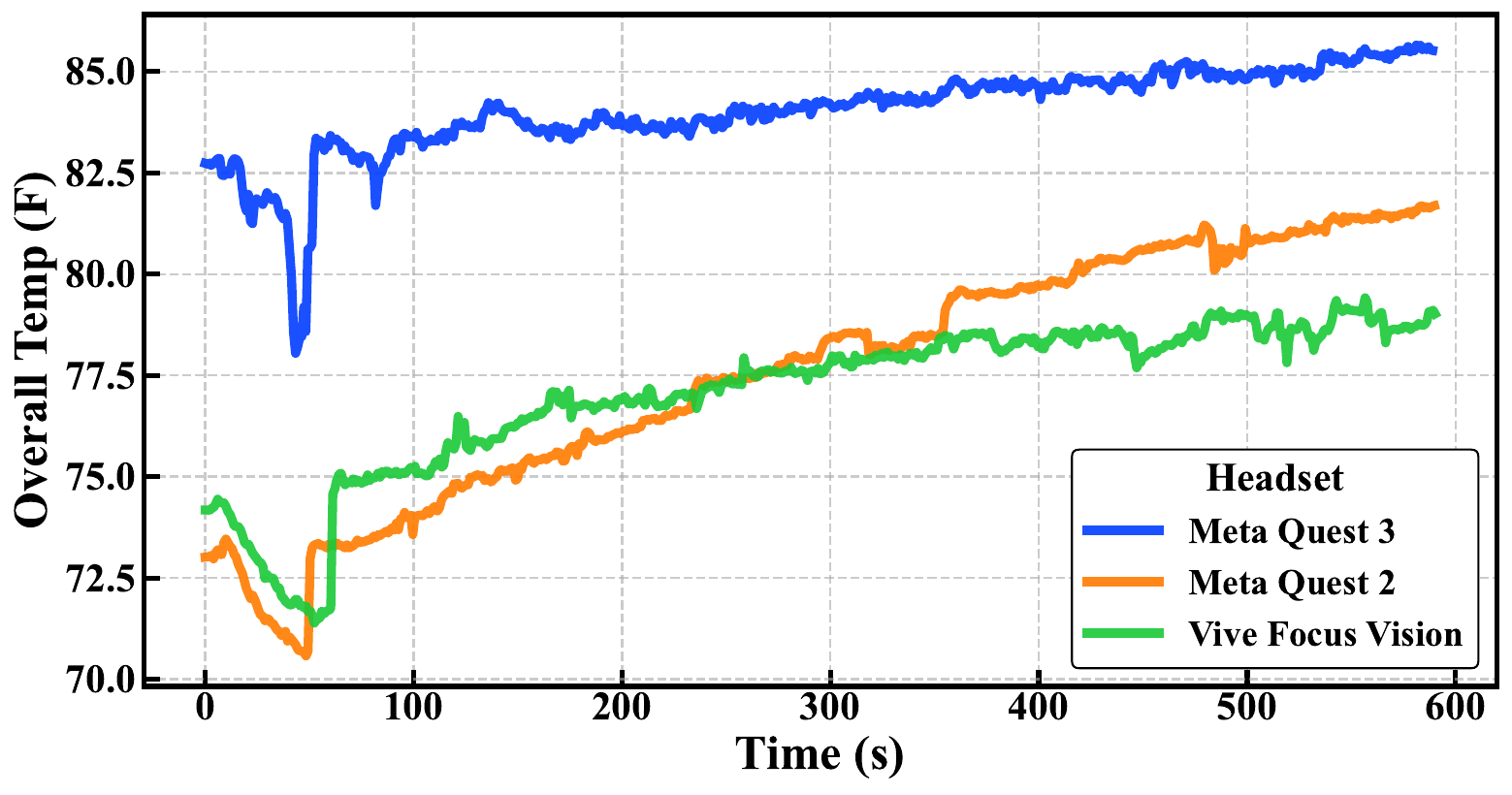}
    \caption{Front-face temperature traces across vs. VR headsets.}
    \label{fig:base_hw_temp_time}
\end{figure}

\noindent\textbf{Hardware Design.} 
Front-mounted batteries concentrate dissipated power on the front of the chassis, which is also the surface most visible to a forward-facing thermal camera. The HTC Vive Focus Vision (one of the devices we evaluated in \cref{subsec:e3_results}) relocates the battery to the back; \cref{fig:base_hw_temp_time} shows that this lowers and flattens the front-face thermal trace under matched workloads. Relocating heat sources away from the externally observable surface does not eliminate the channel, but it materially reduces device specific thermal application profiling. 

\noindent\textbf{Thermal Obfuscation.} Artificially modulating device thermal output (e.g., randomizing fan behavior, injecting dummy workloads) could obscure application signatures. However, this approach incurs power and performance costs.

\noindent\textbf{Physical Shielding.} Thermal-reflective coatings or shrouds could attenuate external thermal radiation, though at the cost of device aesthetics and potentially user comfort.

\section{Conclusion}
\label{sec:conclusion}

We presented \name, the first non-contact thermal side-channel attack that fingerprints VR applications from externally observable infrared emissions. Our results show that a commodity thermal camera can recover application-level information from a VR headset without software access, cable access, or interaction with the victim device. \name achieves strong performance under controlled indoor conditions and remains partially effective under environmental and cross-device variability, establishing thermal radiation as a practical physical side channel for immersive systems. These findings highlight the need to consider thermal emanations when designing privacy protections for future VR platforms.

\bibliographystyle{IEEEtran}
\bibliography{references}

\appendices
\section{Pilot Study}
\label{app:pilot_study}
The application suite in out pilot study covered a set of representative set of apps including idle, media, utility, and simulation workloads, as summarized in \cref{tab:app_list_feasibility}.

\begin{table}[htb]
    \centering
    \small
    \caption{Application test suite used in pilot study.}
    \label{tab:app_list_feasibility}
    \begin{tabular}{lll}
        \toprule
        \textbf{Application} & \textbf{Category} & \textbf{Load Level} \\
        \midrule
        Home        & --         & Base   \\
        YouTube     & Media      & Low    \\
        YouTube VR  & Media      & Low    \\
        First Steps & Utilities  & Medium \\
        Arkio       & Utilities  & Medium \\
        VRFS        & Simulation & High   \\
        First Hand  & Simulation & High   \\
        \bottomrule
    \end{tabular}
\end{table}

\section{Additional Details on Segmentation}
\label{app:segmentation}

\subsection{Segmentation Model Training}
\label{subsec:segmentation_training}
To train the ROI segmentation model, we adopt a semi-supervised labeling strategy to reduce annotation effort. From the full dataset, a subset of frames is manually annotated with headset masks and split into training, validation, and test sets. The segmentation model is trained on the labeled subset and then applied to the remaining unlabeled frames to generate masks for the full dataset. The SRA-Seg train/val/test split was 80/10/10 and the resulting best model is further utilized to run inference on the rest of the dataset to generate ROI masks.
The final segmentation outputs are filtered using the geometric constraints described in \cref{subsec:roi_segmentation} before feature extraction. See \cref{app:segmentation} for further details.

\noindent\textbf{Semi-Supervised Training}
Our entire dataset comprises approximately 107,000 thermal frames. To minimize manual annotation effort for generating the ground truth for our segmentation model, we adopted a semi-supervised labeling strategy:

\begin{enumerate}
    \item \textbf{Down-sampling:} We uniformly sampled the dataset to one-fourth its original size.
    \item \textbf{Manual Annotation:} We manually labeled 3,084 frames for E1, 1276 frames for E2 and 2385 frames for E3 with precise headset masks, splitting them into training (80\%), validation (10\%), and test (10\%) sets.
    \item \textbf{Model Training:} The SRA-Seg was trained on the labeled subset using binary cross-entropy loss with the Adam optimizer.
    \item \textbf{Inference:} The trained models were applied to segment the remaining unlabeled frames in respective dataset. 
\end{enumerate}

\textbf{Geometric Filtering:} To reject malformed segmentation masks (e.g., masks that erroneously capture head straps or user's face), we apply geometric filtering based on the largest connected component:

\begin{itemize}
    \item \textbf{Aspect Ratio Filter:} Valid headset masks must have width-to-height ratios between 1.3 and 2.8 (headsets are wider than tall; extreme ratios indicate rotation or distortion).
    \item \textbf{Solidity Filter:} Mask solidity (area / convex hull area) must exceed 0.80 to reject irregular shapes.
\end{itemize}

Frames failing these filters are marked with NaN values to preserve temporal sequence integrity while excluding unreliable data from analysis.

\section{Datasets and Preprocessing}
\label{app:datasets_preproc}

\subsection{Data Preprocessing}
\label{app:preprocessing}
We apply the same preprocessing pipeline across all the evaluated models (described in \cref{sec:classification}). To avoid data leakage, preprocessing parameters are learned from the training data and applied unchanged to the test data. Features are first standardized using statistics computed on the training set. We then remove low-variance features below the camera's noise-equivalent temperature difference (NETD) threshold, using \(0.0016\), corresponding to \(40\)~mK. Finally, we apply feature selection using ANOVA F-scores before classification. 

\subsection{Dataset Details}
\label{app:datasets}
In \cref{tab:app_suite_main} we show the application suite utilized for our experiments. We further show our dataset details in \cref{tab:dataset_summary}. Our E3 dataset is collected from three distinct devices and the sessions collected are shown in \cref{tab:exp3_dataset}.

\begin{table}[htb]
\centering
\small
\caption{E3 dataset.}
\label{tab:exp3_dataset}
\begin{tabular}{lc}
\toprule
\textbf{Headset} & \textbf{Sessions} \\
\midrule
Meta Quest 3          & 32 \\
Meta Quest 2          & 32 \\
HTC Vive Focus Vision & 20 \\
\midrule
\textbf{Total} & \textbf{84} \\
\bottomrule
\end{tabular}
\end{table}

\section{Additional Results}
\label{app:e1_results}

\subsection{Activity State Detection}
\cref{tab:gk_perf} shows the first part of our hierarchical models performance, activity state detection model.

\begin{table}[htb]
    \centering
    \small
    \caption{Activity-state detection.}
    \label{tab:gk_perf}
    \begin{tabular}{lccc}
        \toprule
        \textbf{State} & \textbf{Precision \%} & \textbf{Recall \%} & \textbf{F1 \%} \\
        \midrule
        Active        & 88.15 $\pm$ 0.38 & 97.29 $\pm$ 0.19 & 92.49 $\pm$ 0.23 \\
        Idle          & 70.53 $\pm$ 1.80 & 33.18 $\pm$ 1.29 & 45.13 $\pm$ 1.4 \\
        \midrule
        Overall(w)     & 85.27 $\pm$ 0.46 & 86.79 $\pm$ 0.38 & 84.74 $\pm$ 0.47 \\
        \bottomrule
    \end{tabular}
\end{table}

\subsection{Active-app Detection}
\cref{tab:per_class_6class_hier_model} shows the app fingerprinting results of E1. 

\begin{table}[htb]
    \centering
    \small
    \caption{Per-app performance of \name for E1.}
    \label{tab:per_class_6class_hier_model}
    \begin{tabular}{lccc}
        \toprule
        \textbf{Application} & \textbf{Precision} & \textbf{Recall} & \textbf{F1} \\
        \midrule
        Arkio        & 88\% & 88\% & 88\% \\
        \rowcolor{blue!20}
        First Hand   & 100\% & 88\% & 93\% \\
        Media Player & 91\% & 91\% & 91\% \\
        \rowcolor{blue!20}
        VRFS         & 91\% & 100\% & 95\% \\
        \rowcolor{blue!20}
        YouTube      & 100\% & 100\% & 100\% \\
        Zoom Web     & 75\% & 75\% & 75\% \\
        \midrule
        Overall      & 91\% & 91\% & 91\% \\
        \bottomrule
    \end{tabular}
\end{table}

\subsection{Performance w/o activity-state detection}

\Cref{tab:per_class_7class} \name app inference performance that includes Home alongside the six active applications (without running idle-detection step).

\begin{table}[htb]
    \centering
    \small
    \caption{Performance w/o activity-state detection.}
    \label{tab:per_class_7class}
    \begin{tabular}{lccc}
        \toprule
        \textbf{Application} & \textbf{Precision} & \textbf{Recall} & \textbf{F1}  \\
        \midrule
        Arkio        & 80\% & 89\%  & 84\%  \\
        \rowcolor{blue!20}
        First Hand   & 89\% & 100\% & 94\%  \\
        Home         & 67\% & 17\%  & 27\%  \\
        Media Player & 83\% & 91\%  & 87\%  \\
        \rowcolor{blue!20}
        VRFS         & 83\% & 100\% & 91\%  \\
        YouTube      & 77\% & 91\%  & 83\%  \\
        Zoom Web     & 70\% & 88\%  & 78\%  \\
        \midrule
        Overall      & 78\% & 80\% & 76\%  \\
        \bottomrule
    \end{tabular}
\end{table}

\subsection{Additional ML Comparison Results}
\Cref{tab:classifier_comparison} shows \name performance of the evaluated ML classifiers for experiment E1.

\begin{table}[htb]
    \centering
    \small
    \caption{ML model performance comparison.}
    \label{tab:classifier_comparison}
    \begin{tabular}{lccc}
        \toprule
        \textbf{Model} & \textbf{Precision \%} & \textbf{Recall \%} & \textbf{F1 \%} \\
        \midrule
        Random Forest & 91.23 $\pm$ 4.0 & 91.07 $\pm$ 4.0 & 91.05 $\pm$ 4.0\\
        XGBoost       & 96.00 $\pm$ 3.0 & 95.00 $\pm$ 3.0 & 95.00 $\pm$ 3.0 \\
        SVM           & 84.23 $\pm$ 9.0 & 83.93 $\pm$ 6.0 & 83.93 $\pm$ 6.0\\
        \bottomrule
    \end{tabular}
\end{table}


\Cref{tab:e2_initial_results} summarizes the effect of environmental normalization under outdoor conditions. Direct indoor-to-outdoor transfer performs poorly due to substantial domain shift, while adding a small number of outdoor calibration sessions significantly improves classification performance. Environmental normalization provides a smaller but consistent additional improvement across all metrics.

\begin{table}[htb]
    \centering
    \small
    \setlength{\tabcolsep}{3pt}
    \caption{Effect of normalization on outdoor experiments.}
    \label{tab:e2_initial_results}
    \begin{tabular}{llccc}
        \toprule
        \textbf{Setting} & \textbf{Feat.} & \textbf{Prec.} & \textbf{Rec.} & \textbf{F1} \\
        \midrule
        Zero-shot & Raw  & 11.64 $\pm$ 5.22 & 15.62 $\pm$ 6.44 & 12.49 $\pm$ 4.95 \\
        Few-shot  & Raw  & 66.56 $\pm$ 8.44 & 61.39 $\pm$ 7.34 & 60.06 $\pm$ 7.78 \\
        Few-shot  & Norm.& 67.25 $\pm$ 10.15 & 64.44 $\pm$ 8.31 & 62.55 $\pm$ 8.88 \\
        \bottomrule
    \end{tabular}
\end{table}

\begin{figure}
    \centering
    \includegraphics[width=0.9\linewidth]{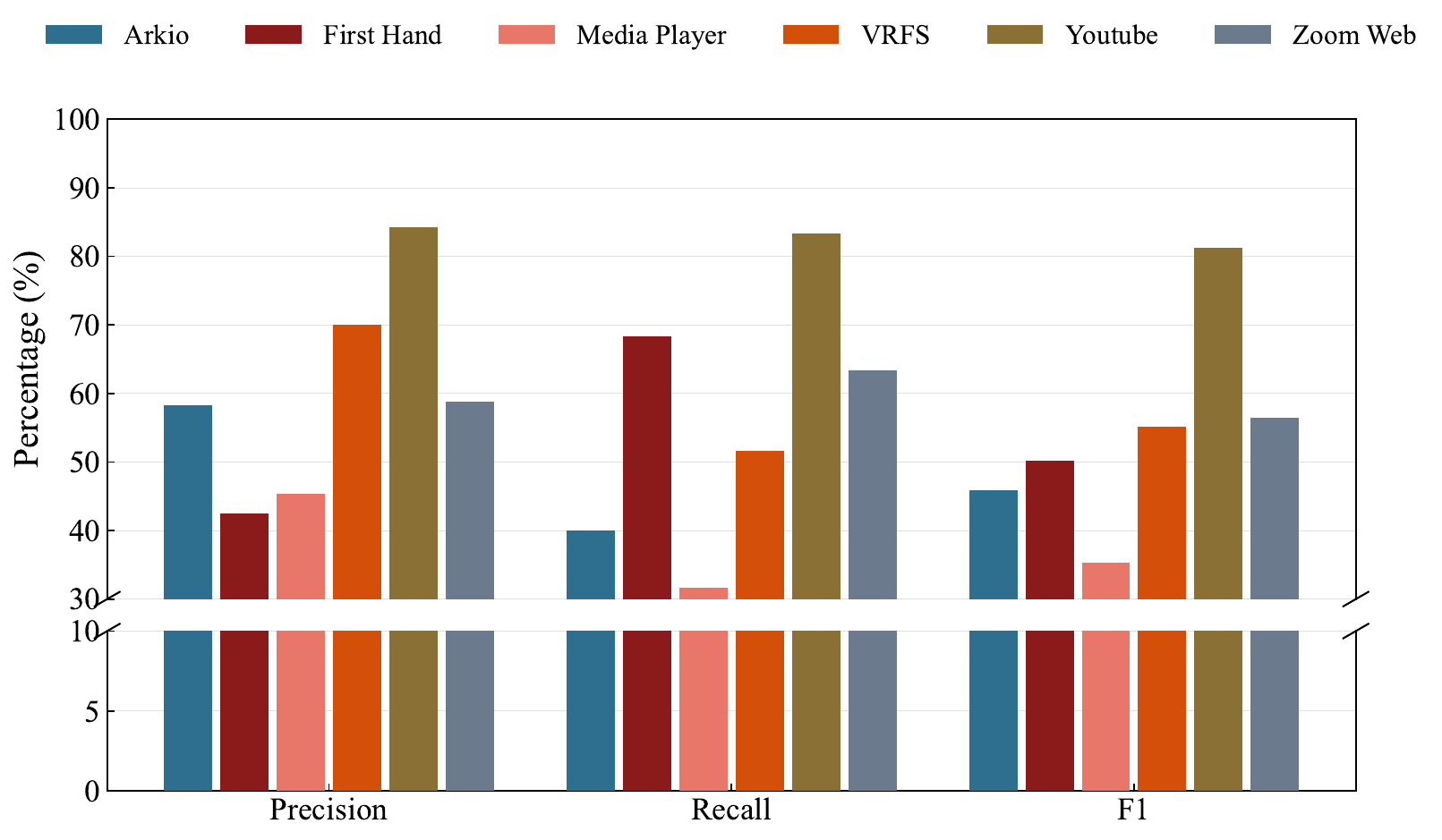}
    \caption{Per-app performance for E2 outdoor experiments (window-level analysis).}
    \label{fig:e2_per_app}
\end{figure}

\Cref{fig:e2_per_app} presents the per-application outdoor fingerprinting performance under the few-shot normalized setting. Applications with more stable and distinctive thermal behavior, such as YouTube and VRFS, remain more separable under outdoor environmental variability than lower-load or less thermally distinctive applications.

\Cref{tab:per_app_metrics_e2} summarizes the per-application outdoor performance in E2 across all folds. The results show that outdoor conditions affect applications unevenly, with YouTube achieving the strongest overall performance while Media Player and Arkio remain more susceptible to environmental variation and inter-class overlap.

\begin{table}[htb]
\centering
\small
\setlength{\tabcolsep}{8pt}
\caption{Mean per-app metrics across all folds for E2.}
\label{tab:per_app_metrics_e2}
\begin{tabular}{lccc}
\toprule
\textbf{App} & \textbf{Precision \%} & \textbf{Recall \%} & \textbf{F1 \%} \\
\midrule
Arkio        & 58.33 $\pm$ 23 & 40.00 $\pm$ 17 & 45.83 $\pm$ 16 \\
First Hand   & 42.45 $\pm$ 9  & 68.33 $\pm$ 22 & 50.22 $\pm$ 9  \\
Media Player & 45.33 $\pm$ 38 & 31.67 $\pm$ 27 & 35.33 $\pm$ 28 \\
VRFS         & 70.08 $\pm$ 21 & 51.67 $\pm$ 25 & 55.20 $\pm$ 15 \\
YouTube      & 84.25 $\pm$ 15 & 83.33 $\pm$ 20 & 81.32 $\pm$ 11 \\
Zoom Web     & 58.81 $\pm$ 23 & 63.33 $\pm$ 26 & 56.49 $\pm$ 16 \\
\bottomrule
\end{tabular}
\end{table}

\begin{table}[htbp]
\centering
\small
\setlength{\tabcolsep}{8pt}
\caption{E2 session-level performance.}
\label{tab:e2_session_per_app}
\begin{tabular}{lccc}
\toprule
\textbf{App Name} & \textbf{Precision \%} & \textbf{Recall \%} & \textbf{F1 \%} \\
\midrule
Arkio        & $68.33 \pm 24$ & $46.67 \pm 16$ & $52.90 \pm 15$ \\
First Hand   & $85.50 \pm 16$ & $81.67 \pm 17$ & $82.32 \pm 13$ \\
Media Player & $66.25 \pm 23$ & $66.67 \pm 18$ & $65.57 \pm 19$ \\
VRFS         & $60.58 \pm 19$ & $61.67 \pm 16$ & $58.76 \pm 13$ \\
YouTube      & $64.56 \pm 19$ & $75.00 \pm 26$ & $66.88 \pm 18$ \\
Zoom Web     & $75.42 \pm 22$ & $63.33 \pm 23$ & $65.04 \pm 17$ \\
\bottomrule
\end{tabular}
\end{table}


\subsection{Diagnostic Headset-Baseline Normalization}
\label{app:headset_norm}

To better understand the role of headset-specific thermal structure in E3, we evaluate a diagnostic headset-baseline normalization strategy. The goal of this analysis is not to introduce a new component of the \name framework, but to test whether removing each headset's static resting thermal profile improves cross-device transfer.

For each headset, we first compute an idle thermal baseline using Home sessions. Let \(B_h(i,j)\) denote the average idle temperature of grid cell \((i,j)\) for headset \(h\). For an active app observation from the same headset, with grid-cell temperature \(T_h(i,j,t)\), we compute the baseline-subtracted residual:

\[
T'_h(i,j,t) = T_h(i,j,t) - B_h(i,j).
\]

This transformation removes persistent thermal offsets associated with the headset's resting state, such as static heat patterns caused by battery location, chassis geometry, material properties, and idle cooling behavior. The resulting residual features represent temperature deviation relative to that headset's idle profile.

However, this normalization only removes a static baseline. It does not account for dynamic device-specific effects, such as differences in fan behavior, heat spreading, thermal inertia, vent placement, or workload-dependent cooling responses. Therefore, even after baseline subtraction, the same application can produce different thermal patterns across headset models.

We also compute a thermal signature ratio to assess whether baseline subtraction increases application-level similarity relative to device-level similarity. Let \(Sim_{app}\) denote the average cosine similarity between the same application observed across different headsets, and let \(Sim_{dev}\) denote the average cosine similarity between different applications observed on the same headset. We define: $R = \frac{Sim_{app}}{Sim_{dev}}.$

A higher value of \(R\) indicates that application-level similarity is stronger relative to headset-level similarity. However, \(R>1\) should be interpreted only as evidence that baseline subtraction increases application-level alignment in feature space; it does not by itself imply that zero-shot classification will be reliable.

The E3 results show that headset-baseline normalization does not solve cross-device transfer. In the pooled setting, performance decreases substantially, indicating that the raw model benefits from device-conditioned application signatures when all headset families are represented during training. In the zero-shot setting, baseline normalization provides only a small improvement and remains far below reliable fingerprinting performance. These results suggest that headset-specific effects are not merely static thermal offsets, but are also embedded in the dynamic way each device dissipates application-induced heat.

\subsection{Additional Ablations}

\begin{figure}[htb]
    \centering
    \includegraphics[width=0.7\linewidth]{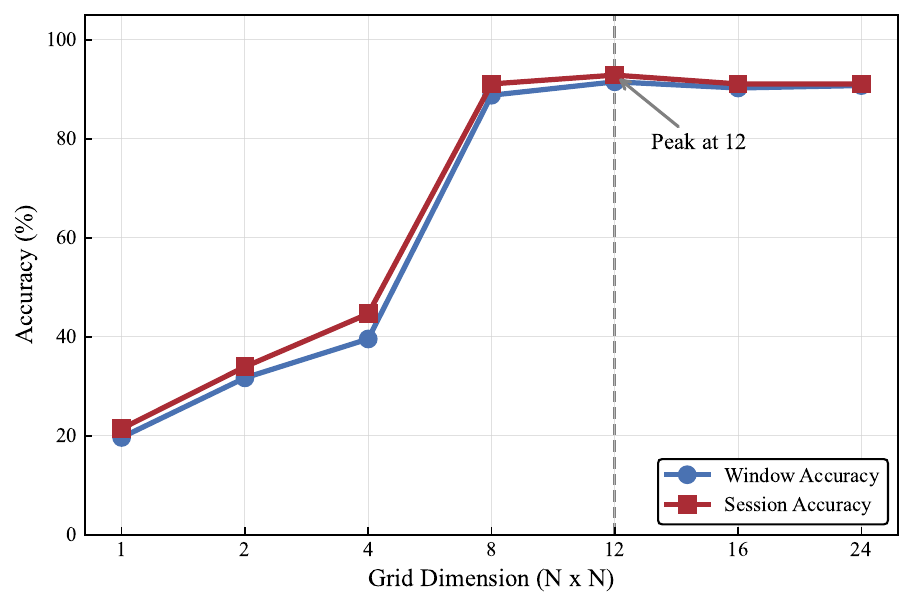}
    \caption{Application classifier accuracy by thermal grid size.}
    \label{fig:grid_performance_plot}
\end{figure}

\Cref{fig:grid_performance_plot} illustrates how thermal grid resolution affects both window-level and session-level classification performance. Classification accuracy improves as the representation preserves more localized thermal structure, peaks near a 12$\times$12 grid, and then saturates or slightly declines at finer resolutions due to increased noise sensitivity.

\begin{figure}[htb]
    \centering
    \includegraphics[width=0.7\linewidth]{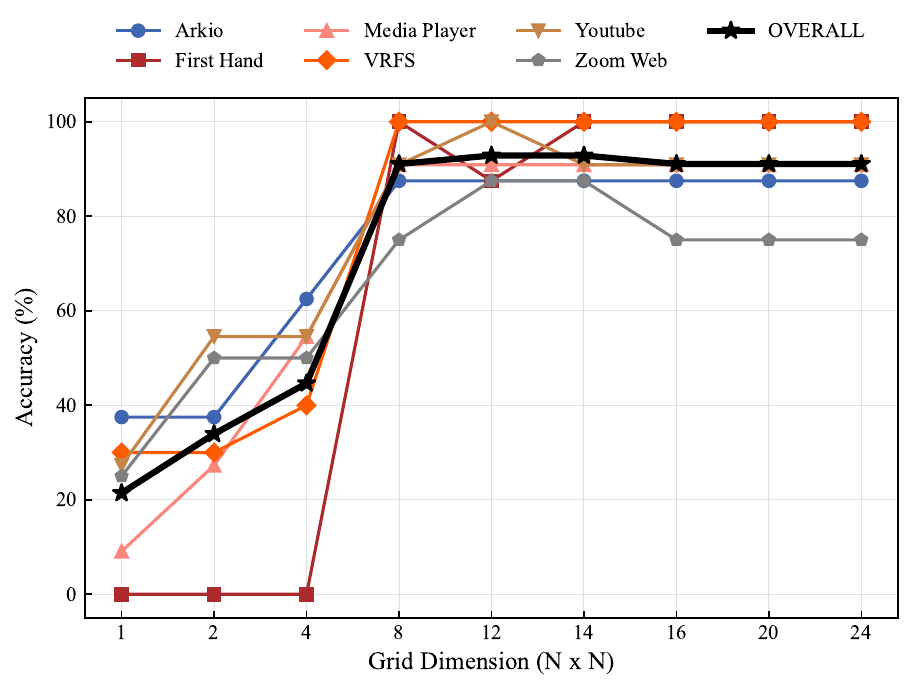}
    \caption{Per app accuracy by thermal grid size.}
    \label{fig:per_class_grid_plot}
\end{figure}

\Cref{fig:per_class_grid_plot} shows the sensitivity of per-application classification accuracy to the spatial grid resolution used in the thermal representation. Most applications benefit from finer spatial grids, although the degree of improvement varies across applications depending on the stability and localization of their thermal signatures.

\begin{figure}[htb]
    \centering
    \includegraphics[width=0.53\linewidth]{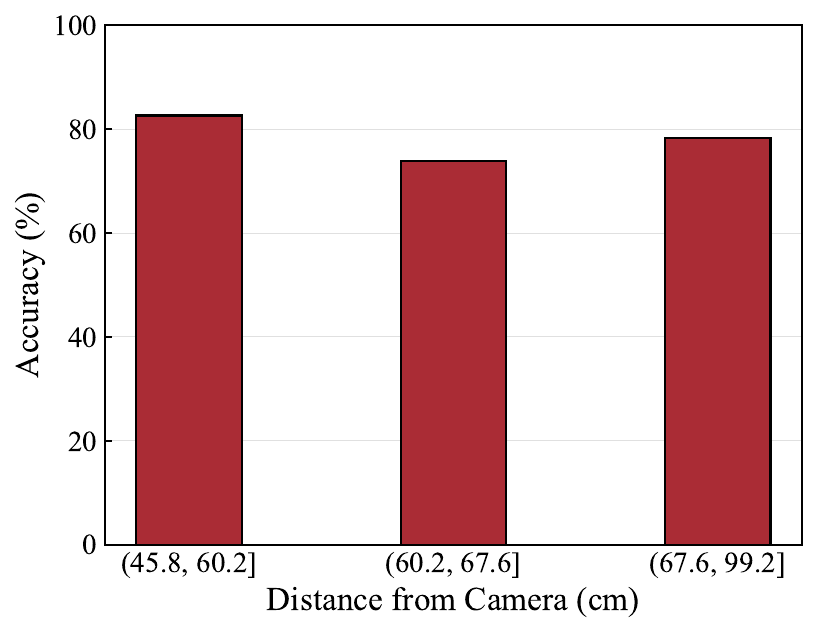}
    \caption{Accuracy vs. distance.}
    \label{fig:dist_acc}
\end{figure}

\Cref{fig:dist_acc} shows how application-classification accuracy varies with camera-to-headset distance. Performance is highest at moderate distances and decreases at larger distances as the observable thermal structure becomes less distinct.

\end{document}